\documentclass[11pt]{article}

\usepackage{amsmath}
\usepackage{amssymb}
\usepackage{dsfont}
\usepackage{epsfig}
\usepackage{graphicx}
\usepackage{hyperref}

\begin{document}
\title{Modeling and Solving Alternative Financial Solutions Seeking}

\author{} \date{}
\maketitle
\begin{center}
Emmanuel Fr\'enod\footnotemark[1]  ~~ Jean-Philippe Gouigoux\footnotemark[2] ~
and ~ Landry Tour\'e\footnotemark[1]{\tiny\hspace{-1pt}'\hspace{1pt}}\footnotemark[2] 

\end{center}
\footnotetext[1]{Universit\'e de Bretagne-Sud,  UMR 6205, LMBA, F-56000 Vannes, France} 
\footnotetext[2]{MGDIS, Parc d'Innovation de  Bretagne Sud, F-56038 Vannes, France}
{
\scriptsize
\noindent {\bf Abstract -} In this paper we model the working of local community finances. As a result of this first step, we
obtain a systemic model that is used to formalize the problem of Alternative Financial Solutions Seeking, which consists in building
a collection of Alternative Multi-Year Prospective Budgets from two Multi-Year Prospective Budgets built by a finance expert.
The modeling and formalization steps are led in a way that allows us to implement a software code for Alternative Financial 
Solutions Seeking based on a Genetic Like Algorithm.
\\
\noindent {\bf Keywords -} Finance Modeling; Local Communities; Systemic Modeling; Optimization; Genetic Algorithms.
}
\newpage
\section{Introduction}
For local communities, political decisions with heavy financial consequences need rigorous 
and detailed studies. 
The purpose of those studies is to provide Decision-Makers with forecasts and projections of their 
financial circumstances to come, for various sequences of projects responding to political goals 
and various ways to finance them.

Tackling those forecasts and projections is a delicate task for experts. Indeed, on the one hand factors 
constraining the projects of a local community are essentially laws, proper management rules which fluctuate
and public opinion which is fickle. On the other hand, the way that those constraints are perceived by 
local community Decision-Makers is also time-varying.

Often, local community calls on experts in order to strike the right balance. Practically, an expert
works in straight collaboration with Decision-Makers of the local community he is engaged with
in order to take into account all the targets and perceived constraints of every project of the local 
community. Its work consists in building financial plans (Prospective Budgets), which are in some sense 
optimized, consistent with the capacity of investment of the local community 
and, of course, comply with the political goals of the Decision-Makers. 
For each prospective budget, the state of 
various indicators, relevant to the resulting financial health of the local community in 
future years, are given. As a result of this work, viable scenarios satisfying partially the political
goals are proposed among which, in an ideal situation, Decision-Makers  may make a choice. 
Unfortunately 
in a large number of non-ideal situations, constraints and goals cannot be satisfied together. 
In those cases the set of viable scenarios influences the evolution of the political goals and the 
constraint perception in order to begin a new iteration of the work process. 

The existing tools dedicated to this iterative work process are somewhat limited. The goal of this paper 
is to set out a new tool to contribute to filling this gap in a specific context we shall describe now.

~

The local communities usually need visibility on their budget over a time period of several years, linked 
to the characteristic duration of political mandate. The main strict constraint, generally imposed 
by current legislation, is that the difference between the receipts and the expenditures cannot be 
negative. That makes-up the balanced budget rule. This balanced budget is in most countries 
shared into sub-budgets which are not necessary balanced. However, each expenditure or 
receipt clearly belongs to a unique sub-budget. For instance, French local communities share 
their budget into an investment budget and an operating budget. 
Positive credit balance amount from the operating budget can be 
transferred to the investment budget. Our work joins in the French model of local 
communities' management but the tackled questioning and the tool we set out are clearly more
general. 

~

In order to explain the goal of the tool we build, we restrict ourselves to the particular case
of a Prospective Budget building where, among the political goals, two objectives are to be reached. 
With software environment avalaible nowadays, a Prospective Budget may be figured out with the first 
objective achieved. In particular, the consequences on the factors involved in the second objective may 
be quantified. Of course, the same can be done exchanging the roles of the two objectives.
Nonetheless, generally speaking, it is not possible to satisfy both objectives, since the constraints are 
too numerous.
Schematically, it may be said that it is possible to bring out two Prospective Budgets $S^1$ and 
$S^2$, where $S^1$ satisfies the first objective, which is symbolized by the  
fact that indicator $V_1$ reaches a targeted value $\widetilde{V_1}$. Prospective Budget $S^2$ satisfies the 
second objective which is translated by $V_2= \widetilde{V_2}$ for a targeted value $\widetilde{V_2}$.
In Prospective Budget $S^1$, $V_2\neq \widetilde{V_2}$ but is determined by the budget building process
which takes constraints into account and which is, in some sense, optimized.
In a similar way, in Prospective Budget $S^2$, $V_1\neq \widetilde {V_1}$.

Having those Prospective Budgets on hand, the next step consists in finding several alternative
ones that are such that neither  $\widetilde{V_1}$ is reached by $V_1$ nor $\widetilde{V_2}$ by $V_2$,
but still satisfy the constraints and are more satisfactory. When this process is executed by an expert, the 
building of those alternative Prospective Budgets uses one more time the tool after having let the targeted values evolve, 
influenced by its knowledge and the interaction with Decision-Makers.  \\

Yet, the new tool which is described in this paper has the ambition of automatically generating a 
collection of alternative 
Prospective Budgets and of introducing them in a usable way, so that Decision-Makers can choose
the one that fits their goals in the best way. \\

In order to create such a new tool, we first identified that the question of finding several 
alternative Prospective Budgets can be formalized as finding a shape, corresponding to the extremum of a given fitness
function, in a multi-dimensional space. 

Then we found out that the best type of optimization methods to tackle this shape search
was Genetic Algorithm. Indeed, Genetic Algorithms have the capability to explore a given domain and, by nature, the result
of a Genetic Algorithm is a set of solutions that optimizes the fitness function.
\label{201310150919} 
For a review on Genetic Algorithms, we refer to  
{Goldberg \cite{citeulike:125978}},
{{Beasley}, {Bull} \& {Martin} \cite{BeasleyBullMartin1,BeasleyBullMartin2}} and
{Davis \cite{davis:handbook}}.

This approach using Genetic Algorithm for an optimization problem is not new.
Nevertheless, the algorithm we propose here has innovative aspects. 
The first one is that we look for an optimal object in a bounded box.
\\
The other innovative aspect is that we look for the argument of an optimum which is not a single point but
a shape in a relatively high-dimensional space.
The use of Genetic Algorithms for shape optimization is classical, and many references exists on the subject.
We refer for instance to {{De Jong} \cite{DeJong}} and {{Castro}, {Ant\`onio} \& {Sousa} \cite{Castro2004356}}. 
We also refer to articles that implement variants of Genetics Algorithms so called Particle Swarm Optimization
(see for instance  {{Mattheck} \& {Burkhardt} \cite{MattheckBurkhardt}} and {{Fourie} \& {Groenwold} \cite{FourieGroenwold}})
and Fuzzy Controlled Genetic Algorithm (see for instance  {{Soh} \& {Yang}  \cite{SohYang}}), both used in structure optimization.
But, in all those references
a Genetic Algorithm is used to drive the successive setting out of the parameters of a software code in order 
to find out the optimal solution. The methods do not use -\,contrary to what we do\,- the capability of Genetic Algorithms 
to directly build a shape in the space. 

Once it was established that Genetic Algorithms is the pertinent tool for our question, it was needed to 
formalize it so as to be able to use Genetic Algorithms.
Yet, the literature concerning them is rich in the context of financial optimization (we refer for instance to 
Chen \cite{ChenSH:2002}).
Besides, the financial modeling is very active on the market finance sector 
(see for instance Goodman \& Stampfli \cite{GoodmanStampfli},  {Ilinski \cite{Ilinski}} and {Fama \cite{Fama1998283}}).
However, it is much less productive for applications in public sector
(see for instance {{Musgrave \cite{Musgrave}} and Rosen \cite{Rosen}}).
Finally, mathematical modeling for finance of local community seems to be very poor (see {Tiebout \cite{Tiebout}}). 

Hence, on the one hand, we had to develop a model of the local community finance system.

On the other hand, we developed a proper formalism (calling upon the model of the local community finance system) 
to develop our Genetic Like Algorithm.
We now summarize this formalism.
It can be considered that any given Prospective Budget is characterized uniquely by the
two values $V_1$ and  $V_2$. In other words, indicators $V_1$ and  $V_2$ become variables
on which Prospective Budget depends. 
To simplify the purpose, 
$V_1$  and $V_2$ are both supposed to be $n$-dimensional, so that it can be assumed that 
$(V_1,V_2)\in{\mathbb{R}}^{2n}$.
Prospective Budget associated with values 
$V_1$ and  $V_2$ writes $S(V_1,V_2)$. Of course, for some values of the variables, say
$(V_1^f,V_2^f)$, Prospective Budget $S(V_1^f,V_2^f)$ does not satisfy the constraints. Then, constraints
may be seen as defining a sub-domain of the space ${\mathbb{R}}^{2n}$ in which the variables lie.
Within this framework, the Prospective Budget  $S^1$ described above writes $S(\widetilde{V_1},V_2^c)$
where $V_2^c$ is computed by the software environment. In the same way, 
Prospective Budget  $S^2$ writes $S(V_1^c,\widetilde{V_2})$.
\\
The method explores, in ${\mathbb{R}}^{2n}$ -- the space where the variables lie, 
the intersection of a box 
containing the two points $(\widetilde{V_1},V_2^c)$ and $(V_1^c,\widetilde{V_2})$, 
associated with the budgets already on hand, and of the sub-domain
where constraints are satisfied in order to identify a shape joining  
$(\widetilde{V_1},V_2^c)$ and $(V_1^c,\widetilde{V_2})$ around which Prospective Budgets
fit well what Decision-Makers are waiting for, are in some sense optimized and satisfy
the constraints.
The box is built by considering in ${\mathbb{R}}^{2n}$ the middle point of
$(\widetilde{V_1},V_2^c)$ and $(V_1^c,\widetilde{V_2})$ 
and by
building in this point an orthonormal frame whose first vector is the normalization of vector
$\overrightarrow{(\widetilde{V_1},V_2^c)(V_1^c,\widetilde{V_2})}$ -- joining 
$(\widetilde{V_1},V_2^c)$ to $(V_1^c,\widetilde{V_2})$.
The other vectors of the frame are exhibited by the mean of the Gram-Schmidt routine.
\\
The method consists in defining a fitness function $F$ which integrates the Decision-Makers' political goals. 
We also have to build the sub-domain on which
budgets satisfy the constraints. 
We define a method to encode the variables in the considered box.  
This coding calls, among others, upon a sub-product of the Gram-Schmidt routine.
Then, we implemented a Genetic Like Algorithm
which consists first in generating a collection of $N$ values $({V_1^l}^0,{V_2^l}^0)_{l=1,\dots,N}$
which are within the box and satisfy the constraints.  
For each value,
Prospective Budget $S({V_1^l}^0,{V_2^l}^0)$ and its fitness 
$F(S({V_1^l}^0,{V_2^l}^0)) = F({V_1^l}^0,{V_2^l}^0)$ can be computed.
By crossover, mutation and constraint management methods, usually combined in Genetic Algorithms, a new collection 
$({V_1^l}^1,{V_2^l}^1)_{l=1,\dots,N}$
(lying in the box and satisfying the constraints)
 is then generated. Going further finally leads to the
$k^\text{th}$ generation $({V_1^l}^k,{V_2^l}^k)_{l=1,\dots,N}$ which may be close to the sought
shape in the intersection of the box  and the sub-domain where constraints are satisfied.
This way of using Genetic Like Algorithm appears to be new.
\\

The main contributions of this paper are the buildings of the models of the local community finance system and of
the Alternative Financial Solutions seeking problem, which
are done in section \ref{201310021035}, and the formalization of those models under a form -\,just evoked\,- allowing
the use of a Genetic Like Algorithm.  
This formalization and the writing of the Algorithm itself are given in section \ref{GLA}.
In the last section, the method is tested, in particular on an operational problem and gives good results. 
This demonstration of this capability of our method is also an important contribution.

\section{Description of the Alternative Financial Solutions seeking problem}
\label{201310021035} 
This section is devoted to the description of the kind of  financial problems we tackle with our method
and tool. We begin by introducing, with a systemic point of view, the French local community yearly budget workings.
Then, we explain the problematic of seeking alternative Multi-Year Prospective Budgets. This is done in section \ref{GLA}.
\subsection{Local community Yearly Budget System working}
\begin{figure}
\begin{center}
\includegraphics[width=0.85\textwidth]{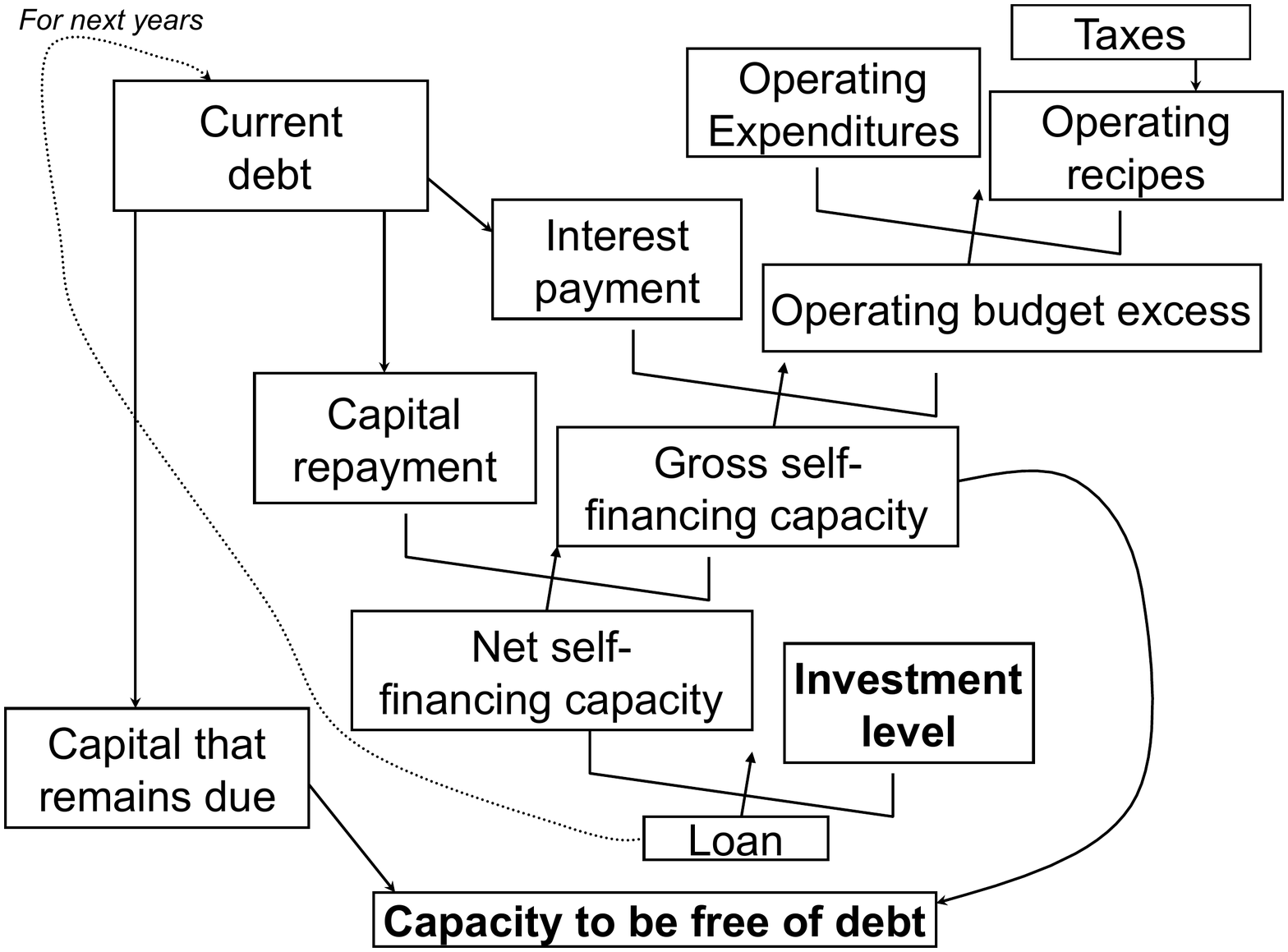}
\caption{French local community Yearly Budget  System working.}
\label{figRepSystBudg} 
\end{center}
\end{figure}
Figure \ref{figRepSystBudg} depicts the schematic working of French local community yearly budget.
To explain this working, we adopt a systemic point of view allowing us to give a global and macroscopic
description, without going into technical or semantical details, of what we call in the following: the Yearly Budget System. 
\\
For readers interested in French local community finance system, we refer to {\cite{QualiteComptable}} and {\cite{PlanM14}}.

~

Among incomes contributing to local community operating budget, there are essentially state allocations
and local "Taxes". Local community cannot influence the state allocation level, however the setting of 
local Tax Level is part of its own competences.    
As a consequence, we consider Taxes as an input of the Yearly Budget System. They lay at the top-right of 
Figure \ref{figRepSystBudg}.

The other inputs of this system are linked with the "Current Debt". They are: the capital associated
to this debt that remains due, the capital that needs to be repaid this year and the interests that
have to be paid. Those amounts are defined by loan contracts of previous years. 
Those inputs are placed on the left-hand side of the figure.
Of course, local community cannot have a direct effect on those inputs, but acts on their values in the
next years by contracting or not new loans. This is symbolized by the dash line in the figure. 
\begin{figure}   
\begin{center}
\includegraphics[width=0.7\textwidth]{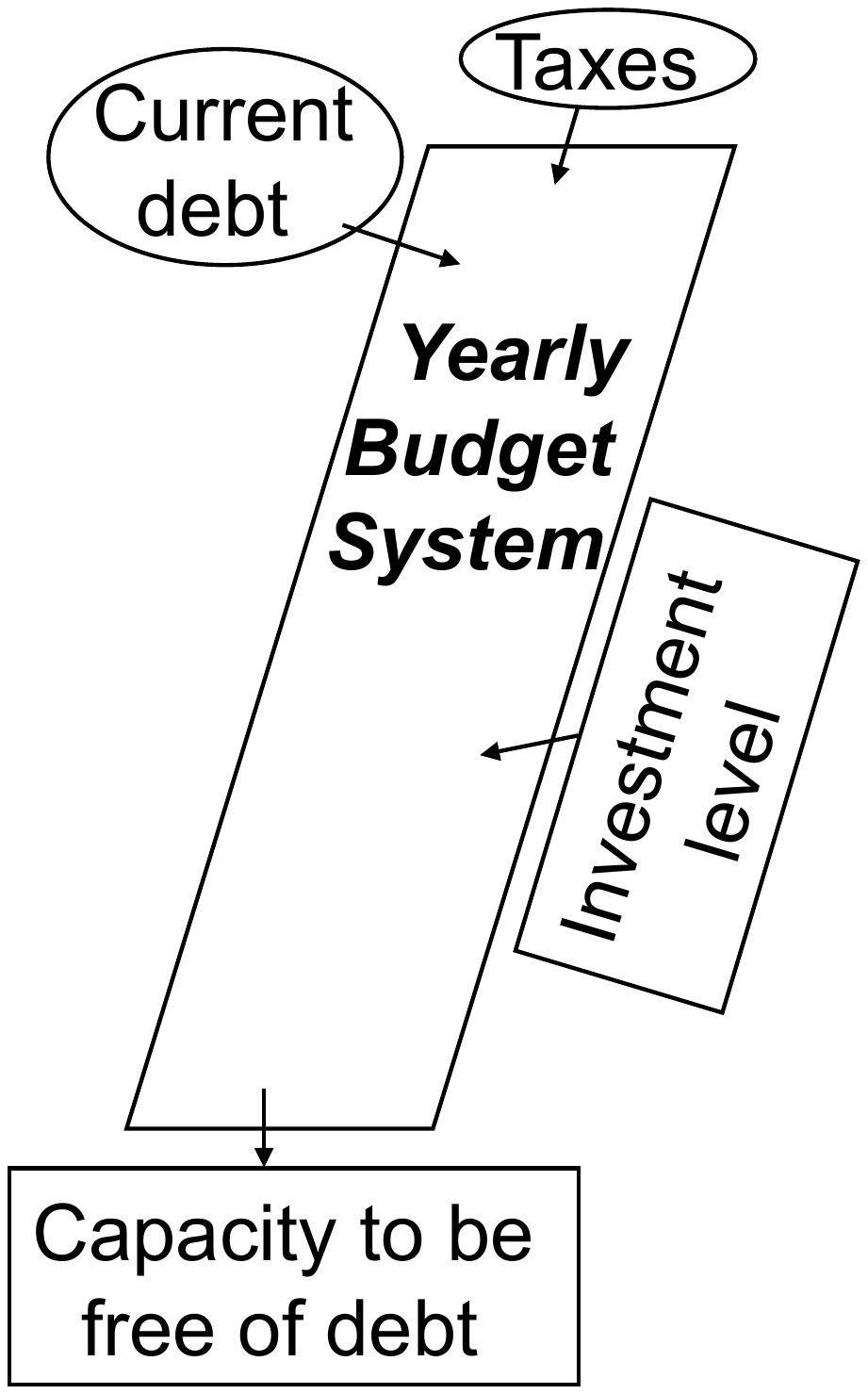}\vspace{-1cm}
\caption{Synthetic Yearly Budget System}
\label{figRepSystSynt} 
\end{center}
\end{figure}
~

Generally, a local community plans to get operating recipes that allow it to face all operating expenditures
and debt interests, and, once those expenditures realized, that leaves a remaining amount that can be used
for investment. This remaining amount is called "Gross Self-Financing Capacity".
This "Gross Self-Financing Capacity" is used to repay the capital that needs to be. The 
remainder, which is called  "Net Self-Financing Capacity", contributes to the investment budget with the
goal to top up subventions and loans to reach the Investment Level wanted by the community. 
This System generates a balanced budget.

At the bottom of the figure, the "Capacity to Be Free of Debt" is mentioned. This indicator is computed
from the capital that remains due and from the Gross Self-Financing Capacity. It is, by definition,
the time (generally expressed in years) for the community to repay all the capital of its debt, if 
no other loans are contracted and if the Gross Self-Financing Capacity remains constant over the 
next years. This indicator is seen here as the output of the system.
A generally-accepted maximum figure for the Capacity to Be Free of Debt is 15 years

~

The Yearly Budget System is presented in Figure \ref{figRepSystSynt} as a synthetic diagram. This diagram
illustrates that Current Debt, Taxes and Investment Level are seen as acting on the Yearly Budget System.
Since Current Debt cannot be influenced directly, only Taxes and Investment Level are considered as
active Inputs of the system. Considering this makes Taxes and Investment Level the variables on which the
Yearly Budget depends and makes the Capacity to Be Free of Debt a result of the Yearly Budget,  or 
in other words, an Output of the system.
In the figure, only three years (\#1, \#2 and \#3) are represented; "\dots" symbolize that other years are coming hereafter.

\begin{figure}
\begin{center}
\includegraphics[width=0.95\textwidth]{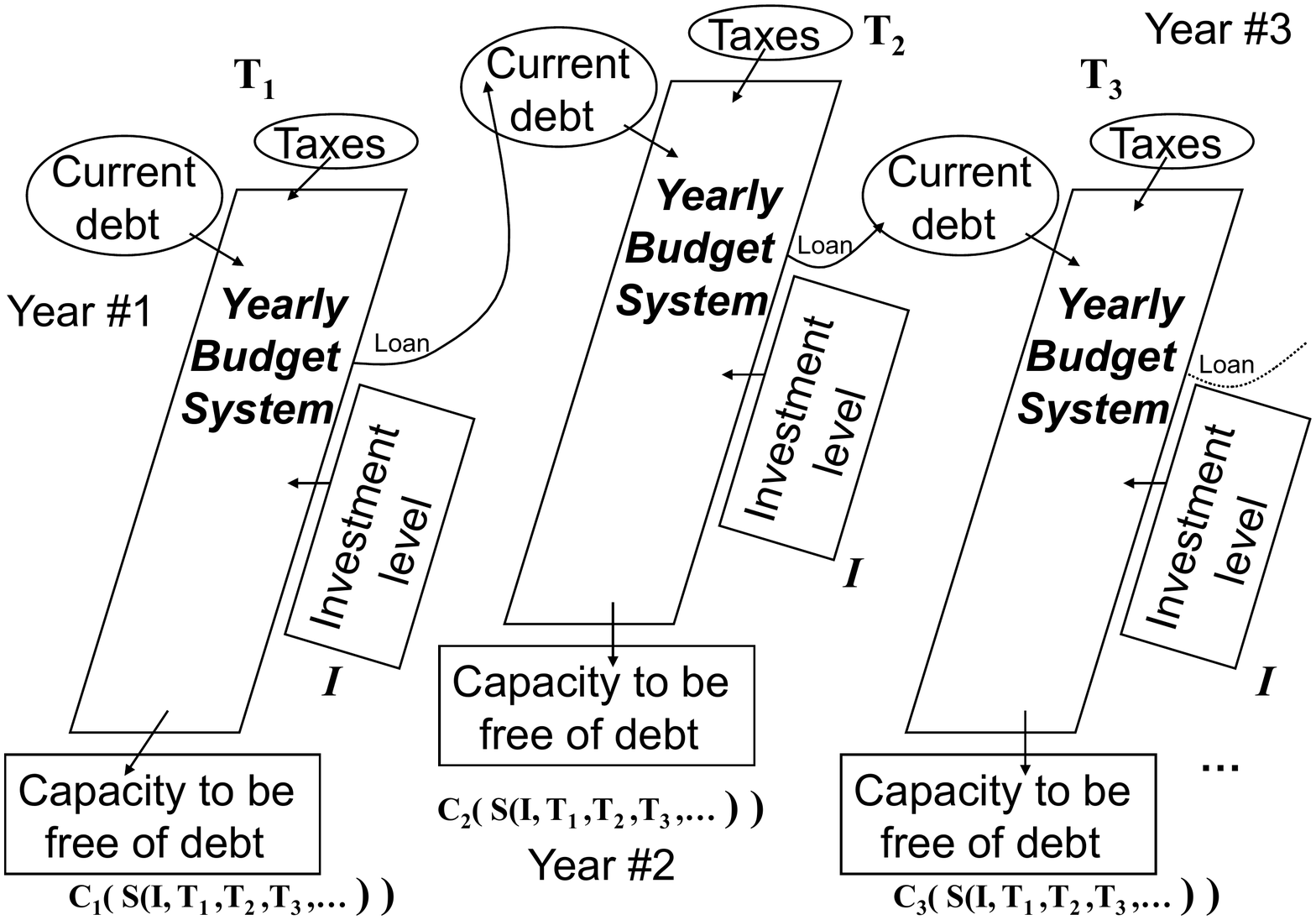}\vspace{-3mm}
\caption{Multi-Year Prospective Budget  System working.}
\label{BudProsMultAnDirect} 
\end{center}\vspace{-5mm}
\end{figure}
\subsection{Multi-Year Prospective Budget Systems}
\label{201310080908} 
From the Yearly Budget System, a multi-year budget may be built. Since those
kinds of multi-year budgets are intended to explore possible futures under several
assumptions, we call them Multi-Year Prospective Budgets. The functioning of such
a Multi-Year Prospective Budget is depicted in Figure \ref{BudProsMultAnDirect}.
On the left-hand side of this picture is drawn the Budget System of the first year. 
This diagram is the synthetic one with arrows coming from the Current Debt box,
 the Tax box and the Investment Level box and an arrow going to the Capacity to 
 Be Free of Debt box.
Loans that are contracted during this first year have a consequence on the debt
of the next years. This is symbolized by the arrow going from the Budget System 
of the first year to the Current Debt box of the second year. Going on, following the 
arrows, it is possible to create a Multi-Year Prospective Budget System for
an arbitrary number of years.

To formalize a bit, the investment leads to the realization of a set of projects, which belong to 
the list of all projects which are desired to be carried out. Then, Investment Level may be described 
using a list of numbers, the cardinal of which being the number of projects. 
Each number indicates if its associated project will be realized or not, and possibly, if it is so, how
close or far from its targeted date it is carried out.\\
Thus, in the case where the Prospective Budget is considered over five years and if five projects are considered, every
possible Multi-Year Prospective Budget depends on ten values 
$(I_1,  I_2, I_3, I_4,I_5,T_1,T_2,T_3,T_4,T_5)=((I_i)_{i=1,...,5},(T_i)_{i=1,...,5})$;  five numbers $(I_i)_{i=1,...,5}$ 
given information on project realization and then indicating the Investment Level, and five Tax Levels:
$(T_i)_{i=1,...,5}$, one for each year. Of course, there are structural constraints on the variables: 
 the $T_i$ cannot be negative.
The Prospective Budget corresponding to a given value 
$(\bar I_1,\bar I_2,\bar I_3,\bar I_4,\bar I_5, \bar T_1,\bar T_2,\bar T_3,\bar T_4,\bar T_5)= ((\bar I_i)_{i=1,...,5},(\bar T_i)_{i=1,...,5})$ of the Investment Level 
and Tax Levels is seen as a solution 
$S(\bar I_1,\bar I_2,\bar I_3,\bar I_4,\bar I_5, \bar T_1,\bar T_2,\bar T_3,\bar T_4,\bar T_5)=S(((\bar I_i)_{i=1,...,5},(\bar T_i)_{i=1,...,5}))$.
The five Capacities to Be Free of Debt are seen as the image associated with a Prospective Budget by a 
mapping. They read:
$(C_k(S(\bar I_1,\bar I_2,\bar I_3,\bar I_4,\bar I_5, \bar T_1,\bar T_2,\bar T_3,\bar T_4,\bar T_5)))_{k=1,...,5} =
 (C_k(S((\bar I_i)_{i=1,...,5},(\bar T_i)_{i=1,...,5}))_{k=1,...,5}$.
\subsection{Alternative Prospective Budget seeking problematic}
Having on hand this formalism, we can insert in it the political goals of Decision-Makers, which are
generally unreachable. At first, we present how part of the political goals may be incorporated to
create two Multi-Year Prospective Budgets which are only partially satisfactory. Then we explain
how, starting from two Budgets, Alternative Prospective Budgets may be sought. Of course, those Budgets
are also only partially satisfactory, but among the generated collection of Prospective Budgets, one
can be preferred to the others and thus chosen among them.
\\

We study the example with five years Prospective Budgets involving five projects, but what is explained in the
following is of course true for any number of years and projects.
In an effort to remain schematic, we consider here that a Political Goal is a given collection
 $(\widetilde I_1,\widetilde I_2,\widetilde I_3,\widetilde I_4,\widetilde I_5,
  \widetilde C_1, \widetilde C_2, \widetilde C_3, \widetilde C_4, \widetilde C_5)=
((\widetilde I_i)_{i=1,...,5},(\widetilde C_i)_{i=1,...,5})$
of targeted Investment Level and targeted Capacities to Be Free of Debt translating into financial terms the projects the 
Decision-Makers plan to get realized and the level of financial sanity they want to reach.
In addition, the Political Goal is provided by a Tax Evolution Pattern, which expresses at what time Decision-Makers
accept tax increases and at what time they prefer stability of tax levels.
 
With the help of a software environment properly programmed, it is possible to compute 
Prospective Budget $S((\widetilde I_i)_{i=1,...,5},(\widetilde T_i)_{i=1,...,5})$, whose every
Yearly Budget is balanced and which satisfies the requested Political Goal, i.e. such that 
$C_k(S((\widetilde I_i)_{i=1,...,5}, $ $ (\widetilde T_i)_{i=1,...,5})) = \widetilde C_k$ for $k=1,...,5$.

However this view is much too naive, since Political Goal $((\widetilde I_i)_{i=1,...,5},(\widetilde C_i)_{i=1,...,5})$ 
generally requires Tax Levels that are incompatible with regulations, or simply not acceptable to
Decision-Makers. 
\\

\begin{figure}[ht]
\begin{center}
\includegraphics[width=0.99\textwidth]{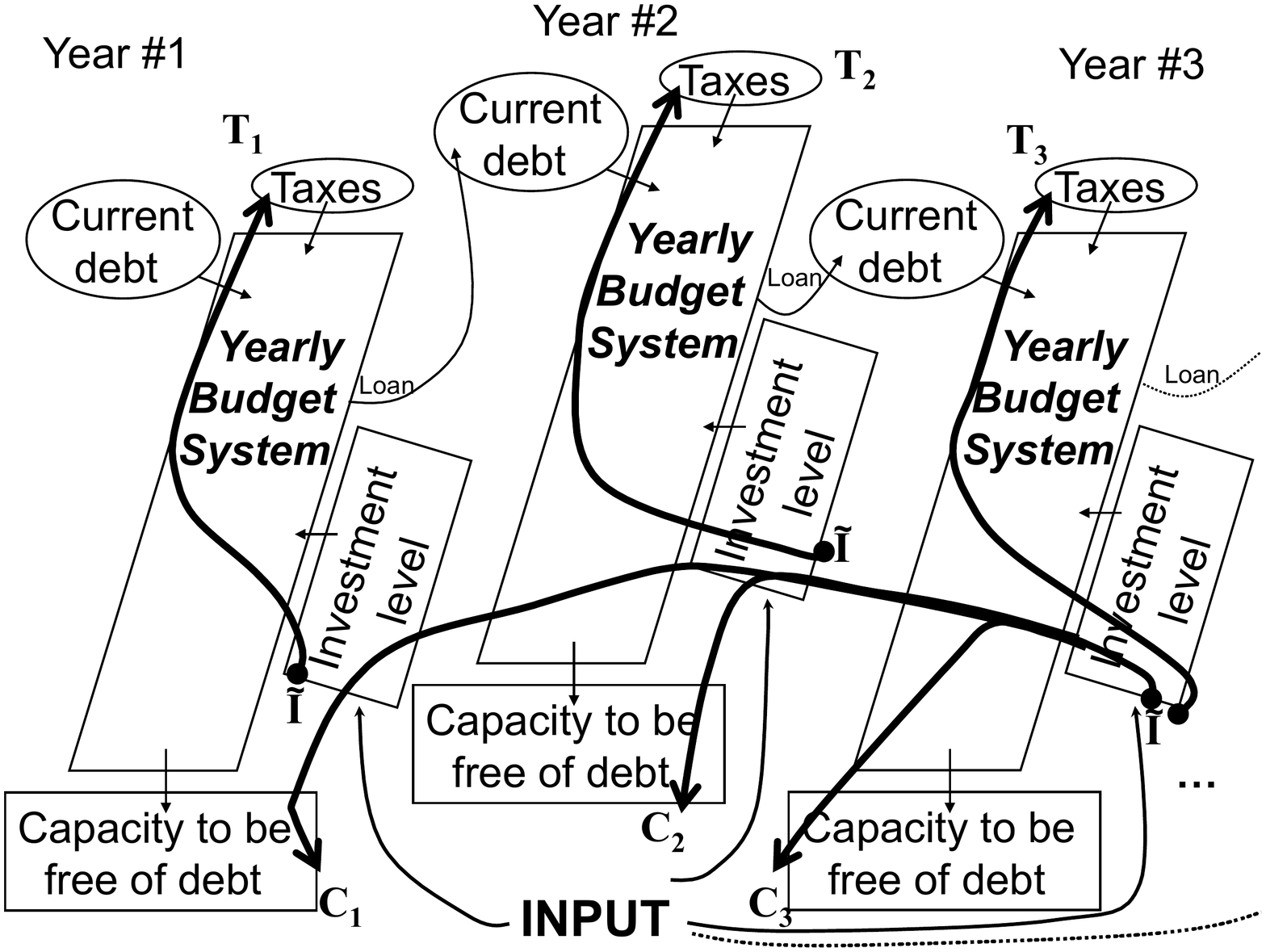}\vspace{-3mm}
\caption{Multi-Year Prospective Budget computed considering Investment Level as input and Tax Levels
and Capacities to Be Free of Debt  as outputs.}
\label{BudProsMultAnInpInv} 
\end{center} \vspace{-5mm}
\end{figure}

The work of Decision-Makers, assisted by public-finance experts, consists in degrading
Political Goal $((\widetilde I_i)_{i=1,...,5},(\widetilde C_i)_{i=1,...,5})$  to be reached with acceptable Taxes Levels. 
This is done with the help of a software environment.  For instance, SOFI software, edited by MGDIS\footnote{\href{http://www.mgdis.fr/}{http://www.mgdis.fr/}},
provides solutions to this question in considering two problems.
Those problems consist, in some sense, in inverting the routine presented in Subsection  \ref{201310080908} and 
Figure \ref{BudProsMultAnDirect}, which was describing how Capacities to be Free of Debt were gotten from chosen Tax and 
Investment Levels.
\\
The first problem, depicted in Figure \ref{BudProsMultAnInpInv}, consists in considering as input the targeted Investment Level  $(\widetilde I_i)_{i=1,...,5}$ and, having
set constraints on Tax Levels, in computing a Multi-Year Prospective Budget
$S((\widetilde I_i)_{i=1,...,5},(T_i^c)_{i=1,...,5})$ such that, for ${k=1,...,5}$, the 
Capacities to Be Free of Debt $C_k(S((\widetilde I_i)_{i=1,...,5},(T_i^c)_{i=1,...,5}))$ 
are as close as possible
(in a given sense) to the Goals $\widetilde C_k$ and with Tax Levels
$(T^c_i)_{i=1,...,5}$ that satisfy the constraints and whose
every Yearly Budget is balanced.
\\
In Figure \ref{BudProsMultAnInpInv}, the diagram of Figure \ref{BudProsMultAnDirect} is used again; and
arrows going from the Investment Level boxes (the investment levels are indicated as inputs) to Tax and
Capacity to Be Free of Debt boxes illustrate that the Taxes and Capacity to Be Free of Debt are computed from the chosen
Investment Levels. 
(For readability, some arrows going from Investment Level boxes to Capacity to Be Free of Debt boxes are not drawn.)
\\

\begin{figure}[ht]
\begin{center}
\includegraphics[width=0.9\textwidth]{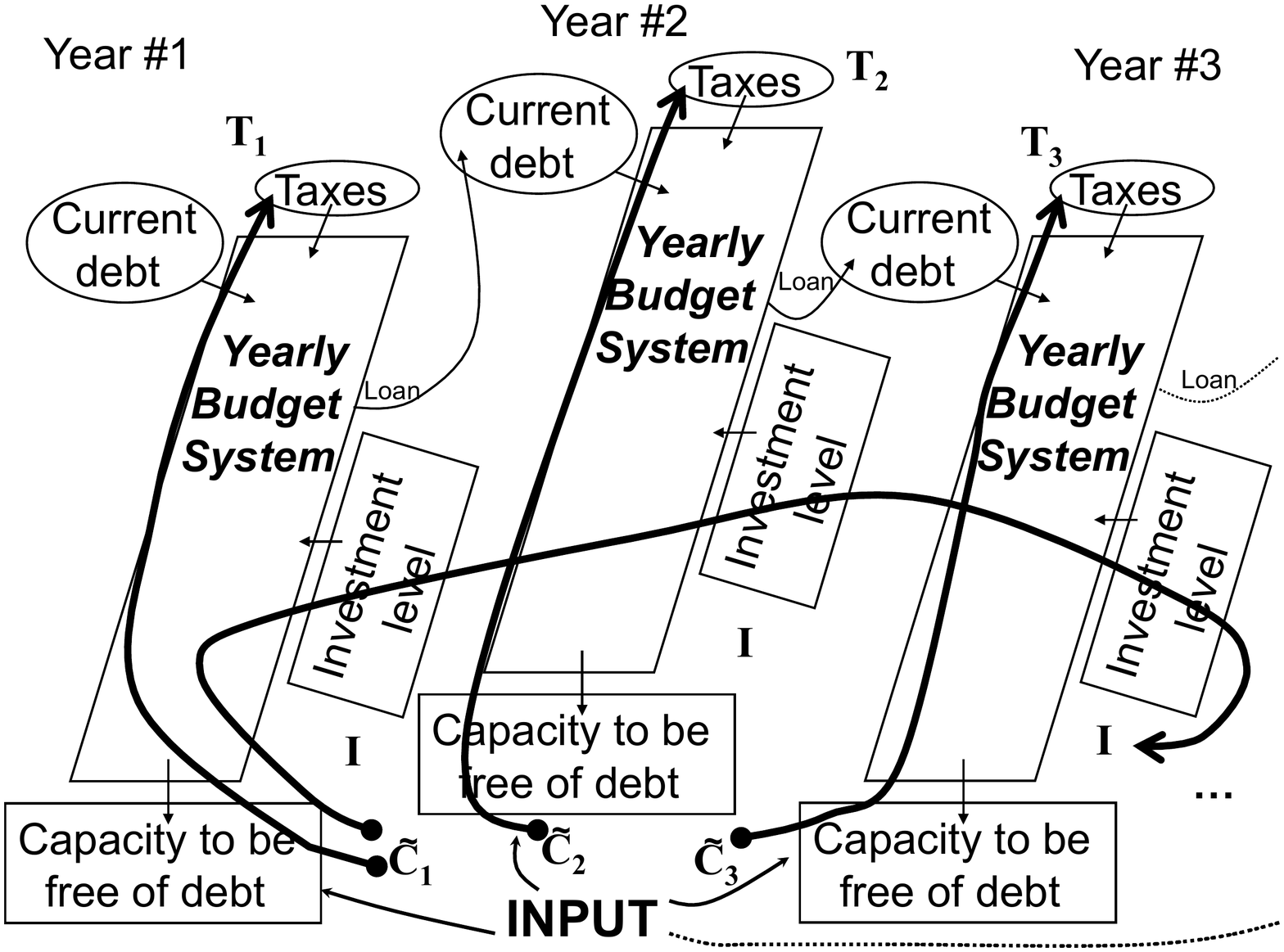} \vspace{-3mm}
\caption{Multi-Year Prospective Budget  computed with Capacities to Be Free of Debt as inputs and Tax Levels
and Investment Levels as outputs.}
\label{BudProsMultAnInpCap} 
\end{center}\vspace{-5mm}
\end{figure}
The second problem we consider is represented in Figure \ref{BudProsMultAnInpCap}, takes the targeted Capacities to Be Free of Debt $(\widetilde C_k)_{k=1,...,5}$ as inputs  and
computes  Multi-Year Prospective Budget  $S((I_i^c)_{i=1,...,5},(\widetilde T_i)_{i=1,...,5})$ 
such that $(\widetilde T_i)_{i=1,...,5}$ satisfies the constraints on Tax Levels,
$C_k(S((I_i^c)_{i=1,...,5},(\widetilde T_i)_{i=1,...,5}) = \widetilde C_k$ for $k=1,...,5$
and
Investment Levels  $(I_i^c)_{i=1,...,5},$ are as close as possible to the Goal $(\widetilde I_i)_{i=1,...,5}$.
\\
In Figure \ref{BudProsMultAnInpCap}, the content of Figure \ref{BudProsMultAnDirect} is used again. However, in this case,
the Capacities to be Free of Debt are indicated as inputs. Arrows going from them to Investment Levels (they are not all
drawn for readability) and to Taxes symbolize that Investment and Taxes are consequences of the chosen Capacity to be 
Free of Debt Levels.
\\

Once those two solutions are set out, they may be evaluated by Decision-Makers, in regards, among others, 
of the Tax Evolution Pattern.

Other Multi-Year Prospective Budgets may then be built by modifying values within
the Political Goal $((\widetilde I_i)_{i=1,...,5},(\widetilde C_i)_{i=1,...,5})$, after financial and
political discussions between Decision-Makers and experts.
This generation of Alternative Financial Solutions or Alternative Multi-Year Prospective Budgets
can be fastidious and long. Thus, only a small number can be generated.

~

The topic of the method and tool we propose here is to automate this generation of 
Alternative Financial Solutions or Alternative Multi-Year Prospective Budgets, and 
to set out a relatively wide number of them, provided with indicators of their quality in order
for Decision-Makers to be able to make a choice between them. 

Roughly speaking, we can consider that Prospective Budgets $S((\widetilde I_i)_{i=1,...,5},(T_i^c)_{i=1,...,5}))$ and
$S((I_i^c)_{i=1,...,5},(\widetilde T_i)_{i=1,...,5}))$ are associated with two points in a 10-dimensional space and that
the possible Alternative Financial Solutions or Alternative Multi-Year Prospective Budgets are gathered
around a geometrical object joining those two points and that have to be sought and identified.
\section{Genetic Like Algorithm}
\label{GLA} 
On the basis of the model we set out in the previous section, we can implement a Genetic Like Algorithm.
Although this approach using Genetic Algorithm is not new in the context of optimization, the algorithm proposed
here has innovative aspects, as explained in the Introduction (see page \pageref{201310150919}). 
Among them they are the fact that we look for an optimal object in a bounded box.
The other innovative aspect is that we look for the optimum not as a single point but
as a shape in a relatively high-dimensional space. For this, we strongly use the fact that the result of a Genetic Algorithm
is a set of solutions that is located on the sought shape.
\\

We go on treating the case when the number of years and the number of projects are both five.
 
\subsection{Dimensionless problem setting}
In order to manage variables and results that are dimensionless and of order one, 
we first rescale the problem. For this purpose, we introduce a characteristic Investment Level describer value
$\textsc{i}$, a characteristic Capacity to Be Free of Debt $\textsc{c}$ and a characteristic Tax Level $\textsc{t}$.
For instance we can chose
\begin{gather}
\label{111} 
\begin{aligned}
 \textsc{i} = &\frac{1}{10} \big(\widetilde I_1 + \widetilde I_2 + \widetilde I_3 + \widetilde I_4 + \widetilde I_5 + 
                                                I_1^c + I_2^c + I_3^c + I_4^c + I_5^c \big)
                    =\frac{1}{10} \sum_{i=1}^{5} \widetilde I_i +  I_i^c, \\
 \textsc{t} = &\frac{1}{10} \big(\widetilde T_1 + \widetilde T_2 + \widetilde T_3 + \widetilde T_4 + \widetilde T_5 + 
                                                T_1^c + T_2^c + T_3^c + T_4^c + T_5^c \big)
                    =\frac{1}{10} \sum_{i=1}^{5} \widetilde T_i +  T_i^c, \\
 \textsc{c} = &\frac{1}{10}  \sum_{k=1}^{5}C_k(S((\widetilde I_i)_{i=1,...,5},(T_i^c)_{i=1,...,5})) + 
                                                                               C_k(S((I_i^c)_{i=1,...,5},,(\widetilde T_i)_{i=1,...,5})),
\end{aligned}
\end{gather}
which are the mean values of values reached by the two Prospective Budgets we have on hand.
Then we define the dimensionless variables and results 
\begin{gather}
\label{222} 
{\cal I}_i  = \frac{I_i }{\textsc{i}}, ~~~ {\cal T}_i = \frac{T_i}{\textsc{t}} \text{ ~ and ~ } \\
\label{223} 
{\cal C}_k({\cal S}(({\cal I}_i)_{i=1,...,5},({\cal T}_i)_{i=1,...,5})) = \frac{C_k({S}(({ \textsc{i}I}_i)_{i=1,...,5},({ \textsc{t} T}_i)_{i=1,...,5}) )}{\textsc{c}}.
\end{gather}

On those variables there are organic constraints:
\begin{gather}
\label{331} 
{\cal C}_k ({\cal S}(({\cal I}_i)_{i=1,...,5},({\cal T}_i)_{i=1,...,5})) \geq 0, ~~~ \text{ ~ for ~ } k=1,...,5,
\end{gather}
which translate the fact that the Capacity to Be Free of Debt is a duration.
There are also constraints linked with legal rules, various regulations and what is politically admissible. These read:
\begin{gather}
\label{330} 
{\cal T}_i \leq {\cal T}_i^\textrm{max}({\cal T}_1,\dots, {\cal T}_{i-1}), \text{ ~ for ~ } i=1,...,5,
\\
\label{332} 
{\cal C}_k({\cal S}(({\cal I}_i)_{i=1,...,5},({\cal C}_i)_{i=1,...,5})) \leq {\cal C}^\textrm{max}, ~~~ \text{ ~ for ~ } k=1,...,5.
\end{gather}
Those constraints, when expressed in dimensionless variables, involve maximum values  
${\cal C}^\textrm{max}$ and $({\cal T}^\textrm{max}_i)_{i=1,...,5}$ which  essentially do not depend on the size
of the concerned local community. 
Inequalities (\ref{332}) express the fact that, at each year, the Capacity to Be Free of Debt  is limited by common rules.
The ${\cal T}^\textrm{max}_i$ in  (\ref{330}) depend on the Tax Levels of the previous years and are 
both imposed by law, which restricts tax evolution, and prescribed by
what community Decision-Makers exclude.

~

\noindent{\bf Remark - }The question of knowing if the fact that every Yearly Budget of every Multi-Year Prospective Budget
needs to be balanced has to enter into the constraint collection may be addressed. The answer is that we work
under the assumption that any Multi-Year Prospective Budget, for instance computed using SOFI, from any variable collection
$(({\cal I}_i)_{i=1,...,5},({\cal T}_i)_{i=1,...,5})$ generates loans and consequently Capacity to Be Free of Debt set 
${\cal C}_k({\cal S}(({\cal I}_i)_{i=1,...,5},({\cal T}_i)_{i=1,...,5}))$
insuring the balanced character of all its Yearly Budgets.

~

Within the dimensionless variables, the dimensionless Political Goals  and other quantities are expressed as
\begin{gather}
\label{333} 
\widetilde{\cal I}_i = \frac{\widetilde I_i}{\textsc{i}}, ~~~ \widetilde{\cal C}_i = \frac{\widetilde C_i}{\textsc{c}}, ~~~
{\cal I}_i^c = \frac{I_i^c}{\textsc{i}} \text{ ~ and ~ } {\cal C}_i^c = \frac{C_i^c}{\textsc{i}},
\end{gather}
and we have on hand two dimensionless Prospective Budgets ${\cal S}((\widetilde {\cal I}_i)_{i=1,...,5},({\cal T}_i^c)_{i=1,...,5}))$ and
${\cal S}(({\cal I}^c_i)_{i=1,...,5},(\widetilde {\cal T}_i)_{i=1,...,5})$, which are associated with two points in a 10-dimensional space,
located not so far from the origin.
\subsection{Fitness choice}
Among the criteria that may enter the fitness definition, they are the targeted values $\widetilde{\cal I}_i$ and $\widetilde{\cal C}_i$
and the Tax Evolution Pattern.

We begin with explaining how easy it is to define a model of Tax Evolution Pattern within dimensionless variables.
In the case where the number of years is 5, it is a collection of 5 non negative values $({\cal A}_k)_{k=1,...,5}$ such that
\begin{gather}                        
\label{444} 
\sum_{k=1}^{5}{\cal A}_k =1,
\end{gather}
and which has the property that ${\cal A}_k = {\cal A}_{k+1}$ in the case of a desired stability of Tax Level between years number $k$
and  number $(k+1)$ and the property that ${\cal A}_k < {\cal A}_{k+1}$ in the case of a planned increase.
Then, a way to measure how far from the Tax Evolution Pattern a given Multi-Year Prospective Budget
${\cal S}(({\cal I}_i)_{i=1,...,5},({\cal T}_i)_{i=1,...,5})$ reduces to computing
\begin{gather}
\label{555} 
F_\textrm{\tiny T} (({\cal I}_i)_{i=1,...,5},({\cal T}_i)_{i=1,...,5}))=
\phi_\textrm{\tiny T}\Bigg(
\sum_{k=1}^{5}\left|
\frac{{\cal T}_k}{\sum_{i=1}^5 {\cal T}_i}
-{\cal A}_k\right|
\Bigg),
\end{gather}
where $\phi_\textrm{\tiny T}$ is a non-increasing function such that $\phi_\textrm{\tiny T}(0)=1$ and 
$\lim_{x\to+\infty}\phi_\textrm{\tiny T}(x)=0$.
In this definition, the division by ${\sum_{i=1}^5 {\cal T}_i}$ allows us to insure
that the values which are compared with the ${\cal A}_k$ range between 0 and 1.

Ways to measure how far from the Politic Goals a Multi-Year Prospective Budget is, are the computations of 
\begin{multline}
\label{666} 
F_\textrm{\tiny I} (({\cal I}_i)_{i=1,...,5})=
\phi_\textrm{\tiny I}\Bigg(\sum_{k=1}^{r}\left| {\cal I}_k-\widetilde{\cal I}_k \right| \Bigg)
\text{  and  } \\
F_\textrm{\tiny C} (({\cal I}_i)_{i=1,...,5},({\cal T}_i)_{i=1,...,5}))=
\phi_\textrm{\tiny C}\Bigg(\sum_{k=1}^{5}\left| {\cal C}_k({\cal S}(({\cal I}_i)_{i=1,...,5},({\cal T}_i)_{i=1,...,5})) -\widetilde{\cal C}_k \right|\Bigg),
\end{multline}
where $\phi_\textrm{\tiny I}$ and $\phi_\textrm{\tiny C}$ have similar definition as $\phi_\textrm{\tiny T}$.

~

With these three functions $F_\textrm{\tiny T}$, $F_\textrm{\tiny I}$ and $F_\textrm{\tiny C}$, 
defining three non negative constants
$\gamma_\textrm{\tiny T}$, $\gamma_\textrm{\tiny I}$ and $\gamma_\textrm{\tiny C}$,
having a sum not too far from 1, we chose the following Fitness Function
\begin{multline}
\label{777} 
F(({\cal I}_i)_{i=1,...,5},({\cal T}_i)_{i=1,...,5}) = \\
\gamma_\textrm{\tiny T} F_\textrm{\tiny T} (({\cal I}_i)_{i=1,...,5},({\cal T}_i)_{i=1,...,5}) + 
\gamma_\textrm{\tiny I} F_\textrm{\tiny I} ({\cal I}_i)_{i=1,...,5}) +
\gamma_\textrm{\tiny C} F_\textrm{\tiny C}  (({\cal I}_i)_{i=1,...,5},({\cal T}_i)_{i=1,...,5}),
\end{multline}
and the largest $F(({\cal I}_i)_{i=1,...,5},({\cal T}_i)_{i=1,...,5}))$ the best the solution
${\cal S}(({\cal I}_i)_{i=1,...,5},({\cal T}_i)_{i=1,...,5}))$.

~

With materials we built, we can reformulate the question of seeking Alternative Financial Solution as follows: we want to exhibit a collection of  $N$ points
$(({{\cal I}^{l}_i}^{\&})_{i=1,...,5},({{\cal T}^{l}_i}^{\&})_{i=1,...,5})$ such that ${\cal S}(({{\cal I}^{l}_i}^{\&})_{i=1,...,5},({{\cal T}^{l}_i}^{\&})_{i=1,...,5})$
satisfied constraints (\ref{330}),
(\ref{331}) and (\ref{332}) and with Fitness worth  $F(({{\cal I}^{l}_i}^{\&})_{i=1,...,5},({{\cal T}^{l}_i}^{\&})_{i=1,...,5})$ as large 
as possible.
\subsection{Frame building by Gram-Schmidt routine}
In the 10-dimensional vector space, for two vectors ${\cal W}=(({\cal J}_i)_{i=1,...,5}),({\cal D}_i)_{i=1,...,5}))$ and 
${\cal W'}=(({\cal J}'_i)_{i=1,...,5}),$ $({\cal D}'_i)_{i=1,...,5}))$,  the following inner
product and norm naturally exist:
\begin{gather}
\label{888} 
\langle{\cal W},{\cal W}' \rangle = \sum_{i=1}^{5}{\cal J}_i {\cal J}'_i + {\cal D}_i {\cal D}'_i 
\text{ ~ and ~ } \|{\cal W} \| = \sqrt{\langle{\cal W},{\cal W} \rangle}.
\end{gather}
Moreover, it  is provided with its canonical basis: 
\begin{gather}
\label{991} 
\begin{aligned}
&{\mathbf e}_1 =((1,0,0,0,0), (0,0,0,0,0)), & &{\mathbf e}_2 =((0,1,0,0,0),(0,0,0,0,0)),
\\
&{\mathbf e}_3 =((0,0,1,0,0),(0,0,0,0,0)), ~ \dots, & &{\mathbf e}_{10} =((0,0,0,0,0),(0,0,0,0,1)).
\end{aligned}
\end{gather}

On the one hand, from the  points $((\widetilde {\cal I}_i)_{i=1,...,5}, ({\cal T}_i^c)_{i=1,...,5})$ and
$(({\cal I}_i^c)_{i=1,...,5},(\widetilde {\cal T}_i)_{i=1,...,5})$, we can build the first
vector of the frame by normalizing the vector linking those two points. This vector is:
\begin{gather}
\label{999} 
{\mathbf g}_1 = \frac{\breve {\mathbf g}_1}{\| \breve {\mathbf g}_1\|},
\text{ ~ where ~ }
\breve {\mathbf g}_1= ((\widetilde {\cal I}_i- {\cal I}_i^c )_{i=1,...,5},({\cal T}_i^c -\widetilde {\cal T}_i )_{i=1,...,5}).
\end{gather}
On the other hand, we look for index ${i_\textrm{b}}$ which is such that the absolute value of the inner product of ${\mathbf g}_1$ by 
${\mathbf e}_ {i_\textrm{b}}$ is as large as possible, i.e. such that  
\begin{gather}
\label{AA1} 
\langle {\mathbf g}_1,{\mathbf e}_ {i_\textrm{b}}\rangle = \max_{i=1,...,10} \{ \langle {\mathbf g}_1,{\mathbf e}_ {i}\rangle \}.
\end{gather}
The basis is then built by induction:
Once $j$ orthonormal vectors are obtained, the $(j+1)^\text{th}$ is gotten by removing from 
${\mathbf e}_ {(i_\textrm{b}+j \mod 10)}$
its projection onto every vector of the new basis already computed and by renormalization, or in other words, by computing
\begin{gather}
\label{AAA} 
{\mathbf g}_{j+1} = \frac{\breve {\mathbf g}_{j+1}}{\| \breve {\mathbf g}_{j+1}\|},
\text{ ~ where ~ }
\breve {\mathbf g}_{j+1} =  {\mathbf e}_ {\eta(i_\textrm{b}+j)}   
  -  \sum_{p=1}^{j}
    \langle {\mathbf e}_ {\eta(i_\textrm{b}+j)},   {\mathbf g}_{p}\rangle {\mathbf g}_{p},
\end{gather}
where $\eta(i) = i$ if $1\leq i \leq 10$ and  $\eta(i) = i-10$ if $10\leq i \leq 20$.

Once all the $({\mathbf g}_{j})_{j=1,...,10}$ are obtained they make an orthonormal basis of the vector space whose first
vector is born by the straight line linking the two points associated with the two dimensionless
Prospective Budgets we have on hand.

With the help of this basis, we will build the box in which we will seek the targeted geometrical object and the coding
of Prospective Budgets.

~

Let ${\mathbf B} $ be the $10\times 10$ matrix such that if ${\cal W}=(({\cal J}_i)_{i=1,...,5},({\cal D}_i)_{i=1,...,5})$ is a vector
expressed in the canonical frame
\begin{gather}
\label{BBB} 
{\cal U} = ({\cal U}_1, \dots,  {\cal U}_{10})= {\mathbf B}{\cal W},
\end{gather} 
gives its coordinates within frame $({\mathbf g}_{j})_{i=1,...,10}$.
The $i^\text{th}$ column of ${\mathbf B}$ is made of the coordinates of ${\mathbf e}_{i}$ within the new frame
and the  $i^\text{th}$ column of ${\mathbf B}^{-1}={\mathbf B}^T$ is made of the coordinates of ${\mathbf g}_{i}$ within the canonical frame. 
In the following we will consider that ${\mathbf B}$ and its inverse matrix ${\mathbf B} ^{-1}$ are known. 
\subsection{Box building and coding}
The geometrical object will be looked after within a box containing the  points
$((\widetilde {\cal I}_i)_{i=1,...,5},$ $({\cal T}_i^c)_{i=1,...,5}))$ and
$(({\cal I}^c_i)_{i=1,...,5},(\widetilde {\cal T}_i)_{i=1,...,5}))$
associated with the two dimensionless Prospective Budgets we have on hand.

The box we chose is the cube centered in the middle point ${\cal M}$  of
$[((\widetilde{\cal I}_i)_{i=1,...,5},({\cal T}_i^c)_{i=1,...,5})),$
$(({\cal I}^c_i)_{i=1,...,5},(\widetilde {\cal T}_i)_{i=1,...,5}))]$, whose coordinates are
\begin{gather}
\label{CCC} 
{\cal M} = \Bigg(\bigg(\frac{{\cal I}^c_i+\widetilde {\cal I}_i}{2} \bigg)_{i=1,...,5},\bigg(\frac{{\cal T}_i^c+\widetilde {\cal T}_i}{2} \bigg)_{i=1,...,5} \Bigg),
\end{gather}
with edges being  $\{2\|\breve{\mathbf g}_1\| {\mathbf g}_1,(\|\breve{\mathbf g}_1\| {\mathbf g}_i)_{i=2,...,10}\}$, where $\breve{\mathbf g}_1$ is defined 
by formula (\ref{999}).\\
Two opposite faces of this hypercubic box are orthogonal to the straight line linking
the points 
$((\widetilde {\cal I}_i)_{i=1,...,5}),$ $({\cal T}_i^c)_{i=1,...,5}))$ and
$(({\cal I}^c_i)_{i=1,...,5}),(\widetilde {\cal T}_i)_{i=1,...,5}))$.

Another (and more usable) way to characterize the box is to say that it is the range of 
$[-1,1]\times[-\frac12,\frac12]^{9}$ by the mapping 
\begin{gather}
\label{DDD} 
{\cal P} \mapsto {\cal M} +\|\breve{\mathbf g}_1\| \, {\mathbf B} ^{-1} {\cal P},
\end{gather}
whose inverse is
\begin{gather}
\label{EEE} 
{\cal R} \mapsto \frac{1}{\|\breve{\mathbf g}_1\|}{\mathbf B}( {\cal R} - {\cal M}),
\end{gather}
where ${\cal M}$ is given by (\ref{CCC}), $\breve{\mathbf g}_1$  by (\ref{999}) and matrix ${\mathbf B}$  
by (\ref{BBB}).

~

Moreover, this transformation gives a coding of any solution ${\cal S}({\cal R})$ by a point in $[-1,1]\times[-\frac12,\frac12]^{9}$.
Hence, without any supplementary effort, we have two codings at our disposal: a solution ${\cal S}({\cal R})$
may be coded by its directly interpretable values $({\cal R}_{1},\dots,{\cal R}_{10})=$ $(({\cal I}_i)_{i=1,...,5}),$ $({\cal T}_i)_{i=1,...,5})$ 
or by the collection of values  $({\cal P}_{1},\dots,{\cal P}_{10})$ that are the coordinates of
point ${\cal P}= ({1}/{\|\breve{\mathbf g}_1\|})$ ${\mathbf B}( {\cal R} - {\cal M}) \in [-1,1]\times[-\frac12,\frac12]^{9}$. 

Generically, in the following we will denote the coding by ${\cal Q} = ({\cal Q}_{1},\dots,{\cal Q}_{10})$.
It will designate the coding by ${\cal P}$ or ${\cal R}$ or any other.

\subsection{Initial Prospective Budget collection}
Fixing the number of members of the collection, and denoting this number by $N$, 
a collection of  $N$ points ${{\cal P}^{l}}^{0}=$
$({{\cal P}_{1}^{l}}^{0},\dots,{{\cal P}_{\hspace{-1pt}10}^ {l}}^{\hspace{-3pt}0})$, for ${l=1,...,N}$,
of  $[-1,1]\times[-\frac12,\frac12]^{9}$ is generated randomly.

The initial collection of solutions is then ${{\cal R}^{l}}^{0}=$
$({{\cal R}_{1}^{l}}^{0},\dots,{{\cal R}_{\hspace{-1pt}10}^ {l}}^{\hspace{-4pt}0})$, for ${l=1,...,N}$,
where  ${{\cal R}^{l}}^{0}= {\cal M} +\|\breve{\mathbf g}_1\| \, {\mathbf B}^{-1} {{\cal P}^{l}}^{0}$.

We assume that every dimensionless Prospective Budget ${{\cal R}^{l}}^{0}$ satisfies all the constraints
(\ref{331}), (\ref{330}) and (\ref{332}). This means that the random generation has to run until 
$N$  dimensionless Prospective Budgets satisfying the constraints are obtained.

Being Generic,  the coding of this initial Prospective Budgets will be denoted by 
${{\cal Q}^{l}}^{0} = ({{\cal Q}_{1}^{l}}^{0},\dots,{{\cal Q}_{10}^{l}}^{\hspace{-4pt}0})$.

\subsection{Constraint management}
For a collection made of  $2N$  individuals,
we will manage constraints by integrating them in the Fitness Function. This consists in adding to Fitness Function
defined by (\ref{777}), the following quantity, or a quantity of this kind,
\begin{multline}
\label{FFF} 
-\phi\Bigg( - \sum_{k=1}^5 \min( {\cal C}_k({\cal S}(({\cal I}_i)_{i=1,...,5},({\cal T}_i)_{i=1,...,5})), 0))\\
+\sum_{k=1}^{5} \max ({\cal C}_k({\cal S}(({\cal I}_i)_{i=1,...,5},({\cal T}_i)_{i=1,...,5})) - {\cal C}^\textrm{max},0 )
+\sum_{k=1}^{5} \max( {\cal T}_k -{\cal T}^\textrm{max}_k,0)
\Bigg),
\end{multline}
for a non-decreasing function $\phi$ such that $\phi(0)=0$ and $\lim_{x\to+\infty}\phi(x) =1$,
multiplied by a factor $\gamma$ relatively large  in front of 1. 
This penalization makes the value of the fitness of Prospective Budgets not respect the constraints to decrease
and then diminishes their chance to pass the selection to come.

\subsection{Algorithm to produce next generation}
Having on hand the $m-$th collection of Prospective Budgets
${\cal S}({{\cal R}^{l}}^{m})$, for ${l=1,...,N}$, and their coding ${{\cal Q}^{l}}^{m}$,
a new collection  ${\cal S}({{\cal R}^{l}}^{m+1})$, with coding ${{\cal Q}^{l}}^{m+1}$
is generated by a usual Genetic Algorithm Like routine we now briefly describe.\\

In a first step, couples of codings of the Prospective Budgets are randomly formed. Then for any couple 
$({{\cal Q}^{l}}^{m},{{\cal Q}^{k}}^{m}) =  
(({{\cal Q}_{1}^{l}}^{m},\dots,{{\cal Q}_{10}^{l}}^{\hspace{-3pt}m}),({{\cal Q}_{1}^{k}}^{m},\dots,{{\cal Q}_{10}^{k}}^{\hspace{-3pt}m}))$,
an integer $i_a$ is randomly chosen among $\{1,2,\dots,10\}$ and the two codings
 $({{\cal Q}_{1}^{l}}^{m},\dots,{{\cal Q}_{i_a-1}^{l}}^{\hspace{-14pt}m\hspace{7pt}}, {{\cal Q}_{i_a}^{k}}^{\hspace{-2pt}m},\dots,{{\cal Q}_{10}^{k}}^{\hspace{-3pt}m})$ 
 and 
 $({{\cal Q}_{1}^{k}}^{m},\dots,{{\cal Q}_{i_a-1}^{k}}^{\hspace{-12pt}m\hspace{5pt}}, {{\cal Q}_{i_a}^{l}}^{\hspace{-2pt}m},\dots,{{\cal Q}_{10}^{l}}^{\hspace{-5pt}m})$
 are generated.
At the end of this first step, we have on hand $2N$ points: the ${{\cal Q}^{l}}^{m}$, for ${l=1,...,N}$ and all the ones generated as described previously
that are denoted ${{\cal Q}^{l}}^{m}$ for ${l=N+1,...,2N}$.\\

The second step consists in making some codings to mutate. For this, a small integer $i_b$, ranging, say, between 0 and $N/50$ 
is randomly generated. Then, $i_b$ codings are randomly chosen among the codings generated in the first step, i.e. among the 
 ${{\cal Q}^{l}}^{m}$ with ${l=N+1,...,2N}$.
For each of them, an integer $i_c$ is randomly chosen among $\{1,2,\dots,10\}$, a number $\nu$ ranging between
$-1$ and $1$ is also randomly generated and the  $i_c-$th component of the concerned coding is incremented by $\nu$.\\

In the third step, the Fitness Function is evaluated on every coding ${{\cal Q}^{l}}^{m}$, for ${l=1,...,2N}$, which results from
the first three steps. In order to do so, it is necessary to determine the ${{\cal R}^{l}}^{m}$, for ${l=1,...,2N}$ associated with the ${{\cal Q}^{l}}^{m}$,
 for ${l=1,...,2N}$,
the Prospective Budgets ${\cal S}({{\cal R}^{l}}^{m})$, for ${l=1,...,2N}$, and finally, the Fitnesses  (penalized by constraints)
$F({{\cal R}^{l}}^{m})$  for ${l=1,...,2N}$.\\

The objective of the fifth step is to randomly select $N$ codings among all the codings ${{\cal Q}^{l}}^{m}$, for ${l=1,...,2N}$ that were brought out by the
previous steps with the principle that the higher the fitness of a coding, the more likely to be selected. In addition, we can use an elitism routine 
which consists in deterministically choosing the $N_\textrm{elit}$ codings that provide the best scores with respect to the Fitness Function.\\

In practice, in order to implement the routine just described, we used the library
"Aforge.Genetics". We conducted a study to measure the impact of the values of "CrossRate" and "MutationRate" parameters
of this library. This study showed that the default values were convenient. Hence we used them.

\subsection{Resizing of obtained solutions}
After several iterations of the algorithm just described, we have on hand a collection of $N$ codings.
${{\cal Q}^{l}}^{\&}$,  for ${l=1,...,N}$. Then, the  ${{\cal R}^{l}}^{\&}=(({{\cal I}^{l}_i}^{\&})_{i=1,...,5},({{\cal T}^{l}_i}^{\&})_{i=1,...,5}))$,  
for ${l=1,...,N}$ are deduced. 
If the coding is the directly interpretable one there is nothing to do. If the coding is based on points ${\cal P}\in[-1,1]\times[-\frac12,\frac12]^{9}$
one needs to apply transformation (\ref{DDD}).

Then, for ${l=1,...,N}$, the resized values $(({{\cal I}^{l}_i}^{\&})_{i=1,...,5},({{T}^{l}_i}^{\&})_{i=1,...,5}))$, 
are computed by inverting formula (\ref{222})
and real Prospective Budgets $S(({{\cal I}^{l}_i}^{\&})_{i=1,...,5},({{T}^{l}_i}^{\&})_{i=1,...,5}))$,   are also computed.
\section{Tests}
The method is now tested on several examples. First, it is tested on one-dimensional problems in order to set out its capability to 
exhibit the point where the maximum of the fitness is located, and, in the case when the fitness shows a plateau with value of 
which is its maximal, to produce a population which is essentially located on the interval which range is the plateau.
Secondly it is tested on the example of local community finances that is described in the previous sections.
\subsection{Test on a one-dimensional problem with a Fitness Function having one maximum}
\begin{figure}[htbp]
\begin{center}
\includegraphics[width=10cm,bb = 1cm 9.5cm 20cm 20cm, clip=true ]{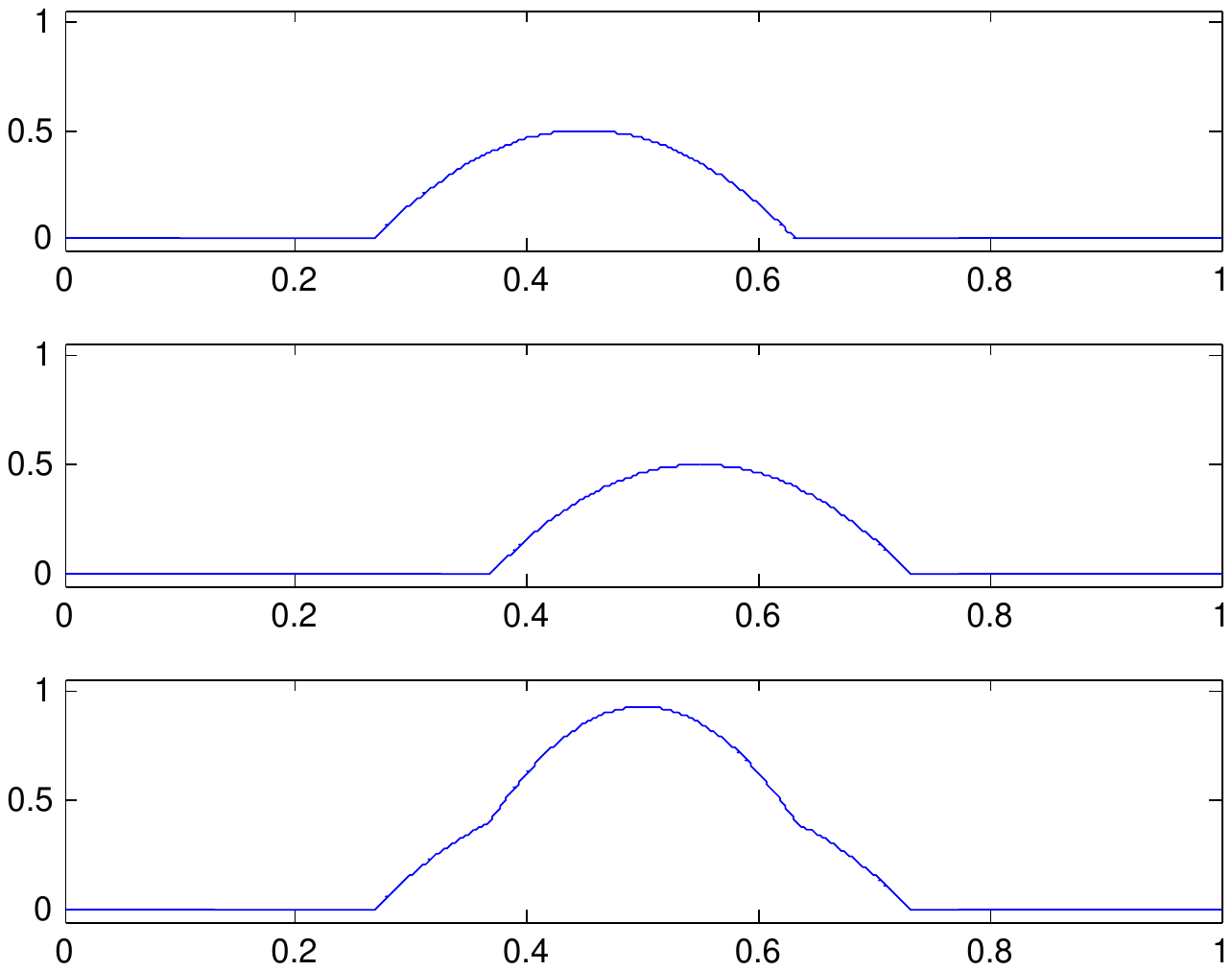}
\caption{Fitness Function (bottom) with one maximum, which is the sum of two functions (top and middle).}
\label{1dWithoutPlateau}
\end{center}
\end{figure}

The Fitness Function considered here is a function with one maximum, which is the sum of
two quadratic ones.
More precisely, defining
\begin{gather}
h_1(x) = \frac12 \max\big(1 -30 \,(x-0.45)^2,0\big), ~~
h_2(x) = \frac12 \max\big(1 -30\, (x-0.55)^2,0\big),
\end{gather}
which are given in Figure \ref{1dWithoutPlateau} at the top and in the middle,
the Fitness Function is
\begin{gather}
F =h_1 +h_2 ,
\end{gather}
defined on $[0,1]$ and
drawn at the bottom of Figure \ref{1dWithoutPlateau}.

We have chosen those functions in order to obtain a function $F$ supported in $(0,1)$ and with values
ranging in $[0,1]$.

A simplified version of the method described above was implemented on this example, with the maxima of $h_1(x)$ and $h_2(x)$
playing roles analogous to those played by
points $((\widetilde {\cal I}_i)_{i=1,...,5}, ({\cal T}_i^c)_{i=1,...,5})$  and $(({\cal I}_i^c)_{i=1,...,5},$ $(\widetilde {\cal T}_i)_{i=1,...,5})$
in the above described method. 

On this example, the method works and gives after 500 generations a collection of points which is very concentrated around $x=0.5$, which
is the argument of the maximum of the fitness function.

Nonetheless its efficiency was compared to optimization methods using a similar  Genetic Like Algorithm, but not involving the two
points around which the argument of the maximum is sought. The method built here was not more efficient.
This seems to lead to the conclusion that the contribution of this method is not to be sought in this direction.

\subsection{Test on a one-dimensional problem with a Fitness  Function having a maximum plateau}
One of the original capabilities of the method described in this paper is that it can give a good representation of the argument of the maximum
of the fitness function when it is an interval.
\begin{figure}[htbp]
\begin{center}
\includegraphics[width=10cm,bb = 1cm 11cm 20cm 19cm, clip=true ]{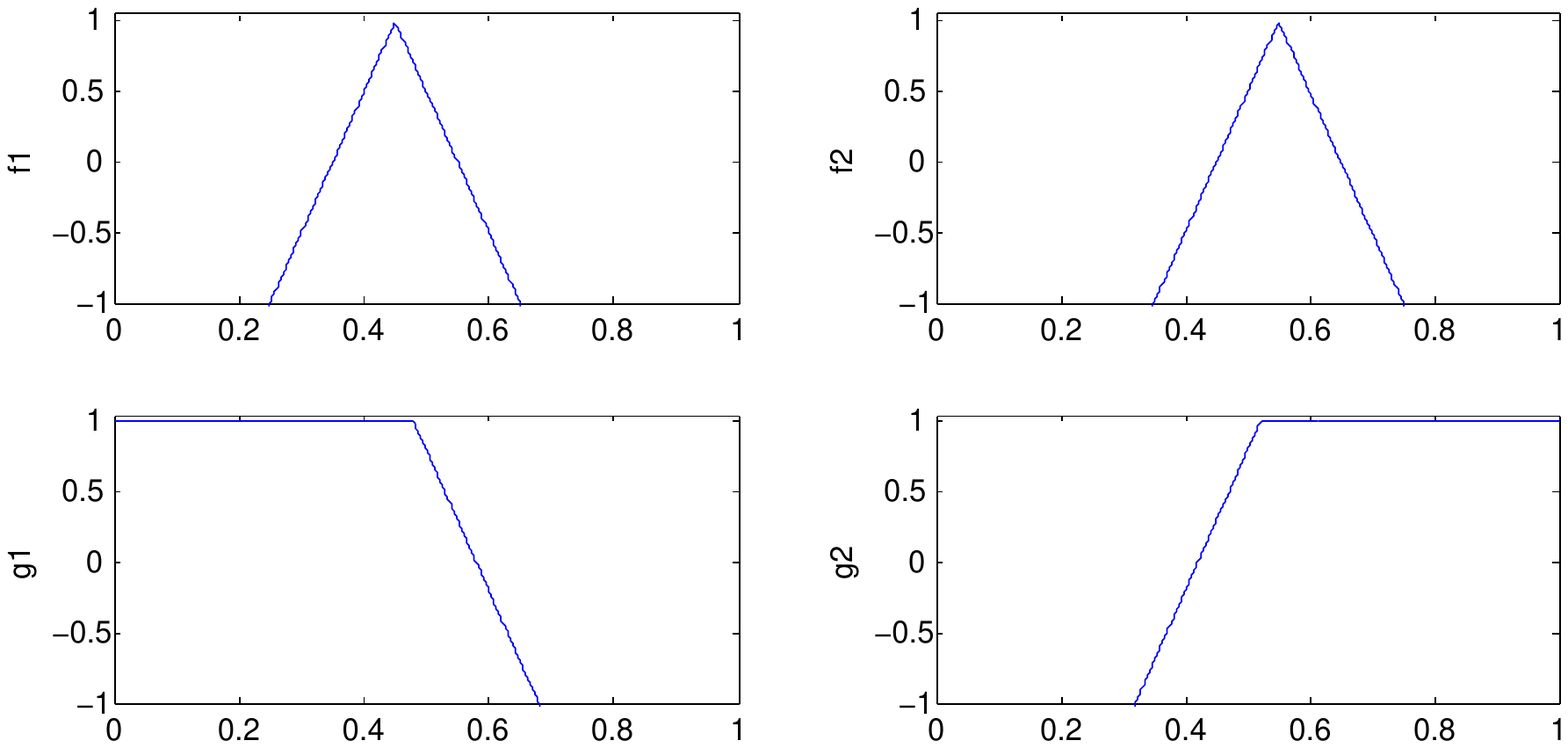}
\caption{functions $f_1$ (top left), $f_2$ (top right), $g_1$ (bottom left) and $g_2$ (bottom right).}
\label{f1f2g1g2}\vspace{-5mm}
\end{center}
\end{figure}
To illustrate this capability, a fitness function having an interval as argument of its maximum will be built.
\\
This fitness function is the sum of two other functions that have both one maximum. 
The result of this sum is a function which 
has a plateau whose span is smaller than the interval defined by the maximum localizations of the two functions it is the sum of.   
\\

\begin{figure}[htbp]
\begin{center}
\includegraphics[width=10cm,bb = 1cm 10.5cm 20cm 19cm, clip=true ]{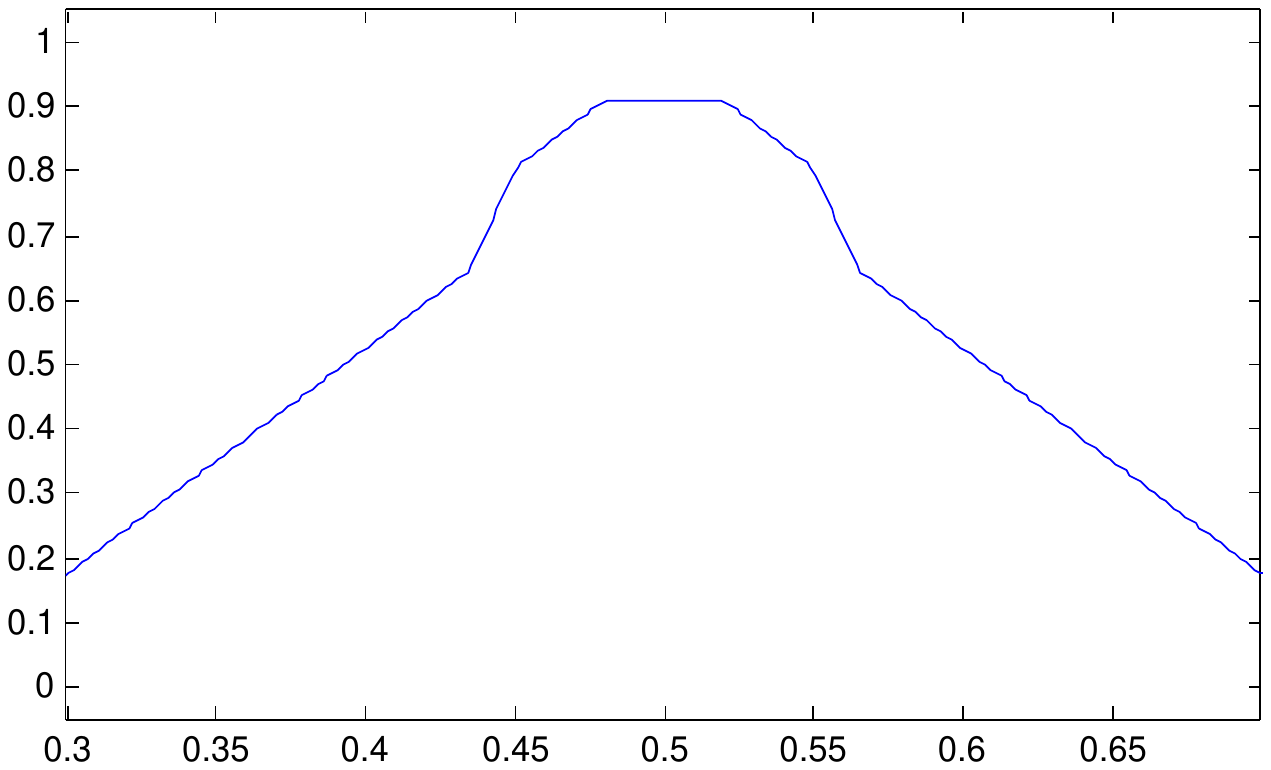}
\caption{Detail of the fitness with  maximum on a plateau which is strictly included in the interval made by the argment-of-the-maximum localizations  
of the two functions it is the sum of.}
\label{1dWithSmallPlateauDetail}
\end{center}
\end{figure}

In practice, in a first place, functions $f_1$ and $f_2$ defined by
\begin{gather}
f_1(x) = 1 -10\, |x-0.45| \text{ and }
f_2(x) = 1 -10\, |x-0.55|,
\end{gather}
 and drawn at the top of Figure \ref{f1f2g1g2} are considered. Functions 
\begin{gather}
g_1(x) = \min\big(1 - 10\,(x-0.48), 1\big) \text{ and }
g_2(x) = \min\big(1 - 10\,(0.52-x), 1\big),
\end{gather}
are also considered.
\begin{figure}[htbp]
\begin{center}
\includegraphics[width=10cm,bb = 1cm 9.5cm 20cm 20cm, clip=true ]{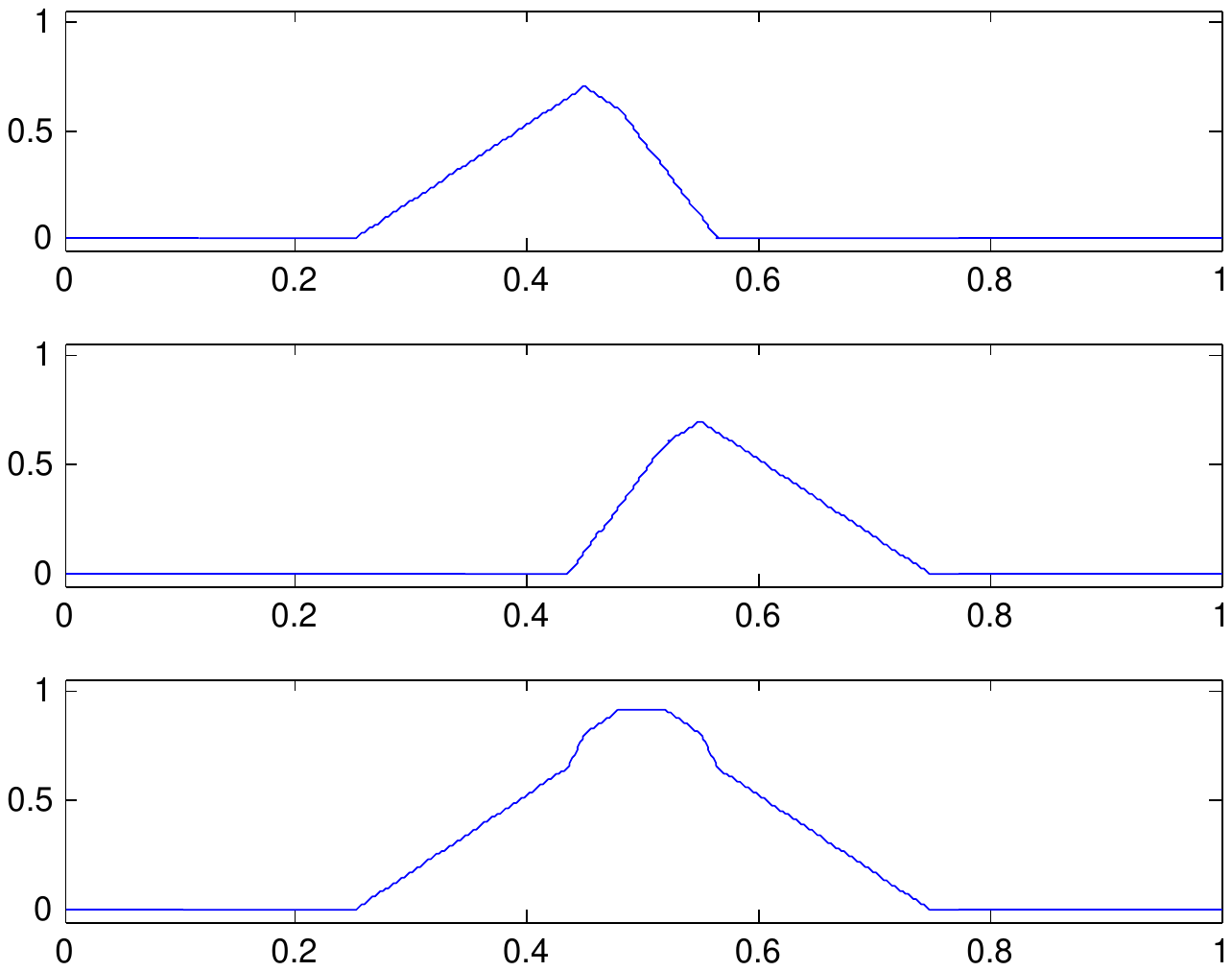}
\caption{The Fitness Function (bottom) with maxima on a plateau which is strictly included in the interval made by the argument-of-the-maximum localizations 
of the two functions (top and middle) it is the sum of.}
\label{1dWithSmallPlateau}
\end{center}
\end{figure}
Finally, the  Fitness Function which is considered is
\begin{gather}
\label{201303292236} 
F = l_1 + l_2.
\end{gather}
It is the sum of the two functions:
\begin{gather}
l_1= 0.7  \max\big(0.5 f_1 +  0.5  g_1, 0\big)  \text{ and }
l_2 = 0.7  \max\big(0.5  f_2 +  0.5  g_2, 0\big).
\end{gather}
Functions $l_1$ and $l_2$ are drawn in Figure \ref{1dWithSmallPlateau} on the top and in the middle;
they both have only one argument of their unique maximum.
Fitness Function $F$ is given in the bottom of this Figure and detailed in  Figure \ref{1dWithSmallPlateauDetail}.
As announced, its maximum makes up a plateau which is the range of an interval ($[0.475, 0.525]$) strictly included 
in the interval ($[0.45, 0.55]$) made by the maximum localizations of the two functions $l_1$ and $l_2$.
\\

A simplified version of the method built in section \ref{GLA} was implemented on this example to locate the arguments of the maximum
of Fitness Function $F$.
In this implemented method, the arguments of the maximum 
of $l_1(x)$ and $l_2(x)$ play the roles that points $((\widetilde {\cal I}_i)_{i=1,...,5}, ({\cal T}_i^c)_{i=1,...,5})$  and $(({\cal I}_i^c)_{i=1,...,5},$ 
$(\widetilde {\cal T}_i)_{i=1,...,5})$ play in the more general method. 
\\
In this simplified version, the collection at each generation is made of 35 points.
The resulting collection  after 500 generations  of the algorithm is given in Table  \ref{tabPlateau}. 
\begin{table}[htdp]
\caption{The collection of points after 500 generations when the Fitness Function is given by \eqref{201303292236}.}
\begin{center}
\begin{tabular}{|c|c|c|c|c|c|c|c|}
\hline\hline
$x$ & 0.489 & 0,491 & 0,491 & 0,500 & 0,492 & 0,489 & 0,487
\\
\hline
$F$ & 0,910 & 0,910 & 0,910 & 0,910 & 0,910 & 0,910 & 0,910
\\
\hline\hline
$x$ & 0,506 & 0,492 & 0,493  & 0,490 & 0,491 & 0,497 & 0,497 
\\
\hline
$F$ & 0,910 & 0,910 & 0,910 & 0,910 & 0,910 & 0,910 & 0,910
\\
\hline\hline
$x$ & 0,494 & 0,493 & 0,489 & 0,496 & 0,489 & 0,494 &1,471
\\
\hline
$F$ & 0,910 & 0,910 & 0,910 & 0,910 & 0,910 & 0,910 & 0,00
\\
\hline\hline
$x$&-0,475 & 0,464 & 0,462 & 0,485 & 0,492 & 0,501 & 0,488 
\\
\hline
$F$ & 0,000 & 0,910 & 0,910 & 0,910 & 0,910 & 0,910 & 0,910 
\\
\hline\hline
$x$ & 0,489 & 0,492 & 0,496 & 0,492 & 0,435 & 0,534 & 0,484
\\
\hline
$F$ & 0,910 & 0,910 & 0,910 & 0,910 & 0,646 & 0,860 & 0,910
\\
\hline\hline
\end{tabular}
\end{center}
\label{tabPlateau} 
\end{table}
The collection is well distributed on the interval ($[0.475, 0.525]$)
whose range is the plateau of Fitness Function $F$. In particular, this final collection does not undergo concentrations that could orientate by mistake interpretation towards concluding that 
the Fitness Function has isolated maxima.
\\

This capability of the method is very important for reaching operational problems. This is what is done in the last test.
\subsection{Test on an operational problem} 
\label{SubsecTOP} 
\begin{figure}[htbp]
~\hspace{-1.3cm}
\includegraphics[width=17.5cm, bb = 2.5cm 17.8cm 18.5cm 27.5cm, clip=true]{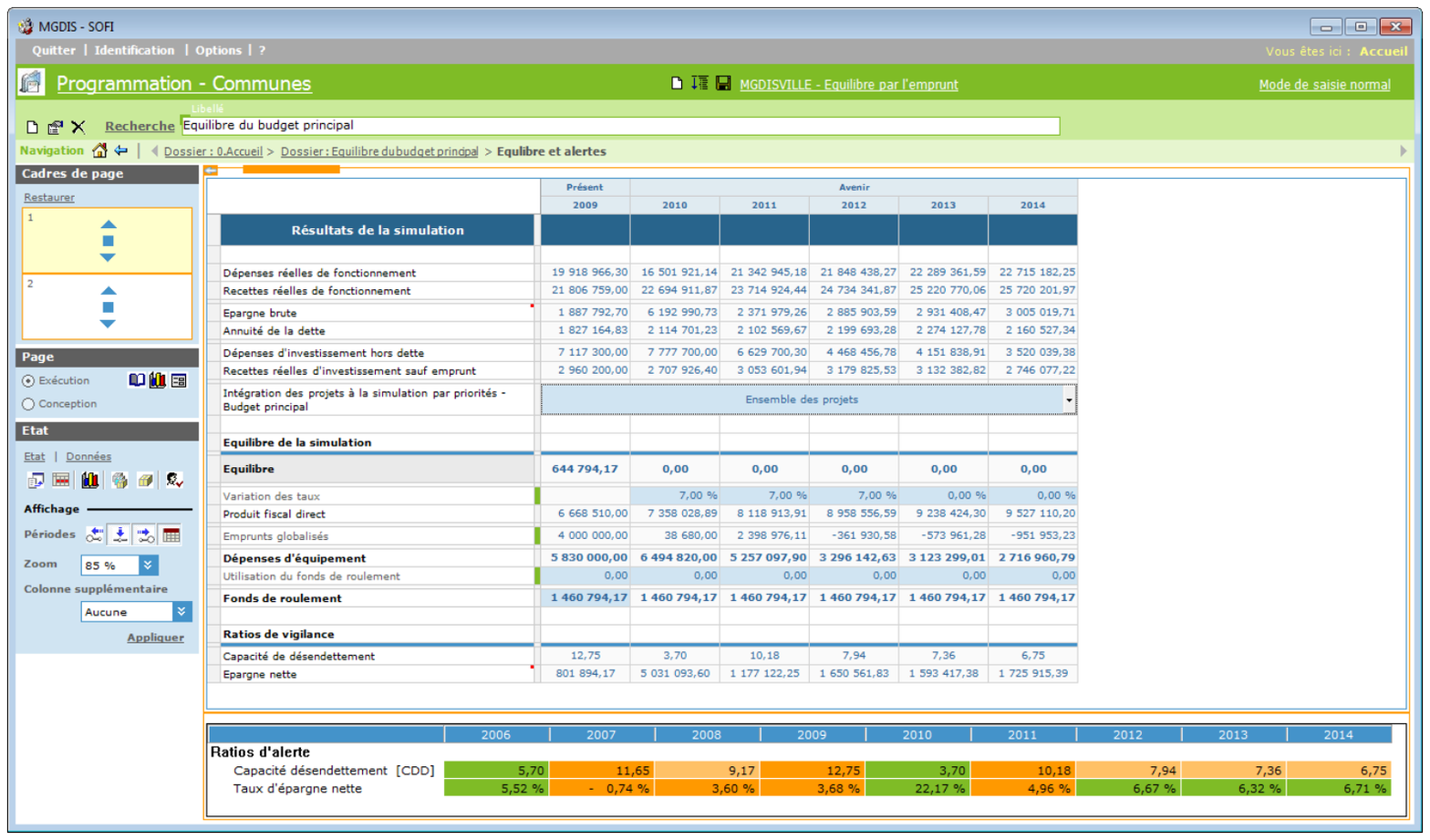}
\caption{The quite liberal solution. (The software has a French interface;  the translations are: Capacit\'e de d\'esendettement = Capacity to Be Free of Debt,
Variation des taux = Tax Increase, Produit fiscal direct = Taxes, D\'epenses r\'eelles de fonctionnement = Operating Expenditures, 
Recettes r\'eelles de fonctionnement = Operating Recipes, Epargne brute = Operating Budget Excess.)
}
\label{LabelFig10D1} 
\end{figure}
\begin{figure}[htbp]
~\hspace{-1.3cm}
\includegraphics[width=17.5cm, bb = 2.5cm 17.8cm 18.5cm 27.5cm, clip=true]{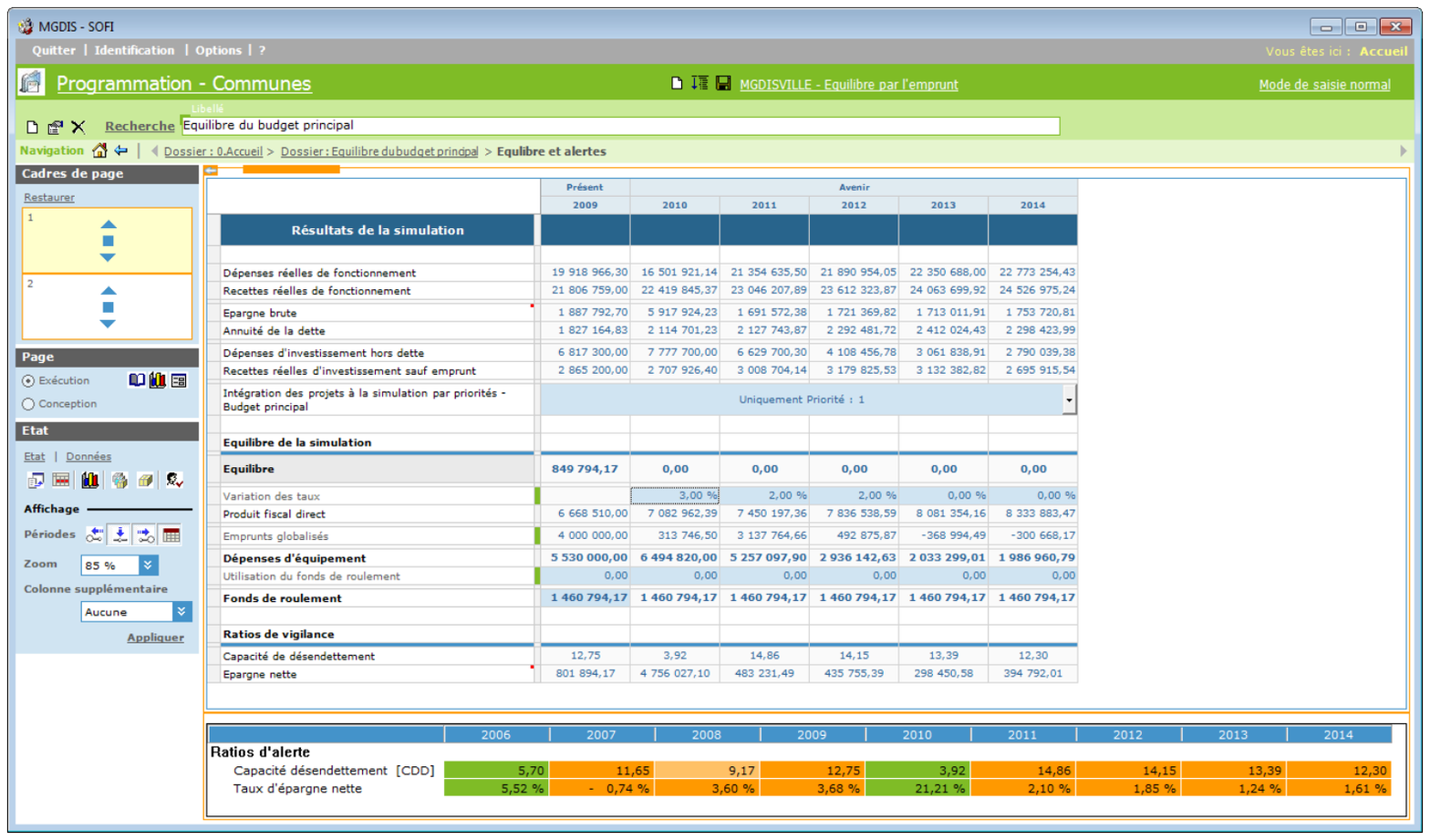}
\caption{The careful solution.}
\label{LabelFig10D2} 
\end{figure}
The method set out in section \ref{GLA} is now tested in a realistic situation.
The experience uses a software product, called SOFI, dedicated to the optimization of local communities' budgets, with a French interface. 
In this subsection, we describe the test and in subsection \ref{SubsecPCCT}, we discuss technical aspects concerning
the way we tuned the method and the way we prevented the appearance of problems we met.
\\

In order to keep the study as realistic as possible, the budget model and numbers that have been chosen are drawn from an actual customer 
of the company MGDIS. The city council (city of which we will not cite the name for confidentiality reasons) used SOFI to simulate the 
impact of a dozen projects onto the budget of the community.
Using this application, a financial expert (Thomas Hody) provided us with two Multi-Year Prospective Budgets.
They correspond to ideal but not reachable situations. The results of our Genetic Algorithm based optimization process have 
been validated by the same person as a solution that indeed improves on the use of the community financial resources.

The first solution, presented in Figure \ref{LabelFig10D1},  is quite liberal, with all projects being realized, and the taxes increased at the maximum rate of 7\% in the first three years
(see the second line of the second part of the Table in Figure \ref{LabelFig10D1}).  
The Capacity to Be Free of Debt ratio remains in the acceptable range (see the "Capacit\'e de d\'esendettement"  line at the bottom of the table in Figure \ref{LabelFig10D1}).\\
The second solution we picked is a much more careful, with only the top priority projects being done, and a very limited increase of tax applied so that the capacity to be free of debt ratio 
remains below 15 years, which is the prudential limit. Figure \ref{LabelFig10D2} shows the values for this second solution.\\

\begin{figure}[htbp]
~\hspace{-1.3cm}
\includegraphics[width=17cm, bb = 2.5cm 19.5cm 18.5cm 27.5cm, clip=true]{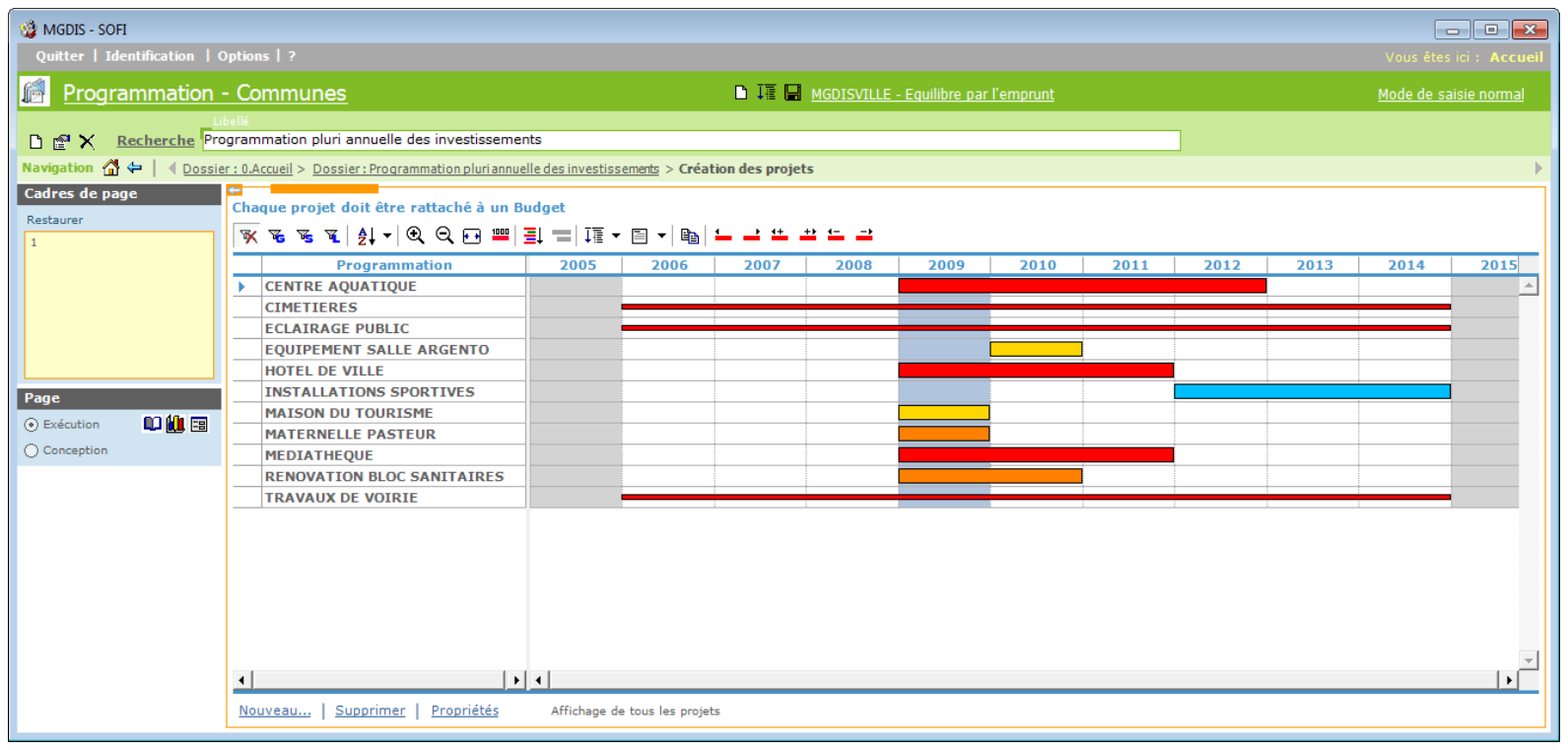}
\caption{The proceedings of the projects for the careful solution.}
\label{LabelFig10D3} 
\end{figure}

Figure  \ref{LabelFig10D3} shows a representation of the proceedings of the projects for the careful solution. A color code helps spotting the priorities of the projects (from hight priority: red, 
orange, yellow, blue: low priority ). It should be noted that, in our example, priority one projects account for the vast majority of the budget.

The coding follows such rules:
\begin{itemize}
\item The first five genes code the five less-than-one-priority projects (the priority-one projects are always active).
\item The coding is equally distributed for each unit: below half, the project is inactive, above half, the project is active.
\item The next five genes contain the evolution of tax, written as a double.
\item We consider the limits to be 0\% at the minimum, and 7\% at the maximum. Thus, the decoding will realize a modulus 0.07 function on the values.
\end{itemize}
The fitness function that was used works as follows:
\begin{itemize}
\item The number of projects brings a linear satisfaction, and accounts for 25\% of the final fitness.
\item The evolution of the tax is best at 0\% and worst at 7\%.
\item Its average evolution accounts for 10\% of the global fitness, and its evolution in the last two years accounts for 5\%.
\item The capacity to be free of debt is optimal at 0 year and worst at 15 years and more. This accounts for 25\% of the final grading of the solution.
\item The capacity to spare money is optimal at 5\%, and accounts for 25\% of the global fitness.
\item No variation at all (in the tax evolution) gives the best results, and this part of the grade accounts for the remaining 10\% of the global fitness.
\end{itemize}
The language used is C\#, and the Genetic Algorithm framework is the one from AForge, which is an Open Source project. The definition of the two vectors corresponding to the solution points are programmed as:
\begin{gather}
\begin{aligned}
 \text {\tt Vector v1 = } & \text {\tt new Vector(new double[] \{ 0.75, 0.75, 0.75, 0.75, 0.75,}
\\
& \text {\tt  0.07, 0.07,  0.07, 0.00, 0.00 \});}
\\
\text{\tt Vector v2 = } & \text {\tt new Vector(new double[] \{ 0.25, 0.25, 0.25, 0.25, 0.25,}
\\
& \text {\tt 0.03,  0.02, 0.02, 0.00, 0.00 \}); }
\end{aligned}
\end{gather}

\begin{figure}[htbp]
~\hspace{-1.3cm}
\includegraphics[width=17.5cm, bb = 2.5cm 17.8cm 18.5cm 27.5cm, clip=true]{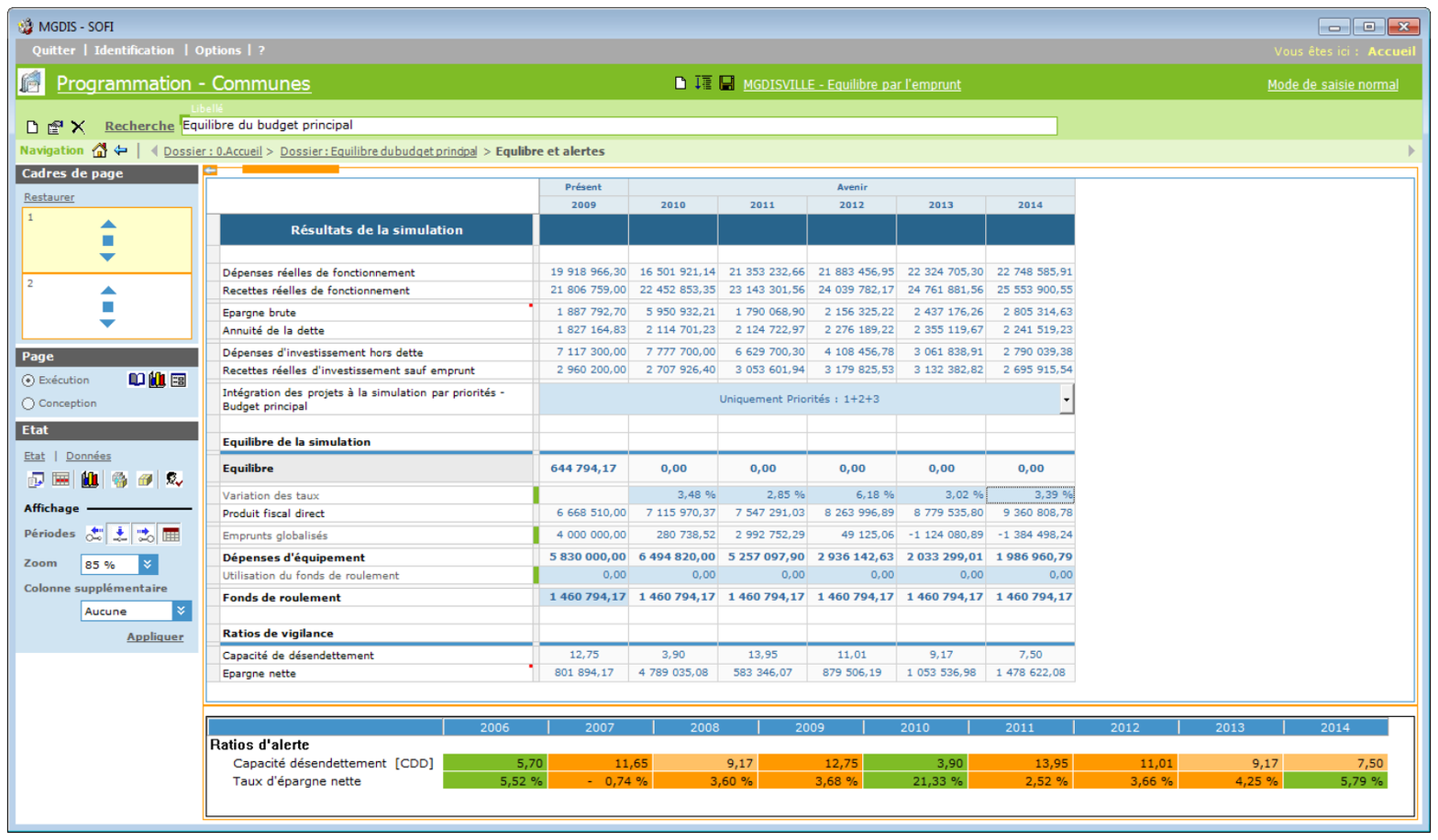}
\caption{The optimal solution.}
\label{LabelFig10D4} 
\end{figure}

\begin{figure}[htbp]
~\hspace{-1.3cm}
\includegraphics[width=17.5cm, bb = 2.5cm 19.8cm 18.5cm 27.5cm, clip=true]{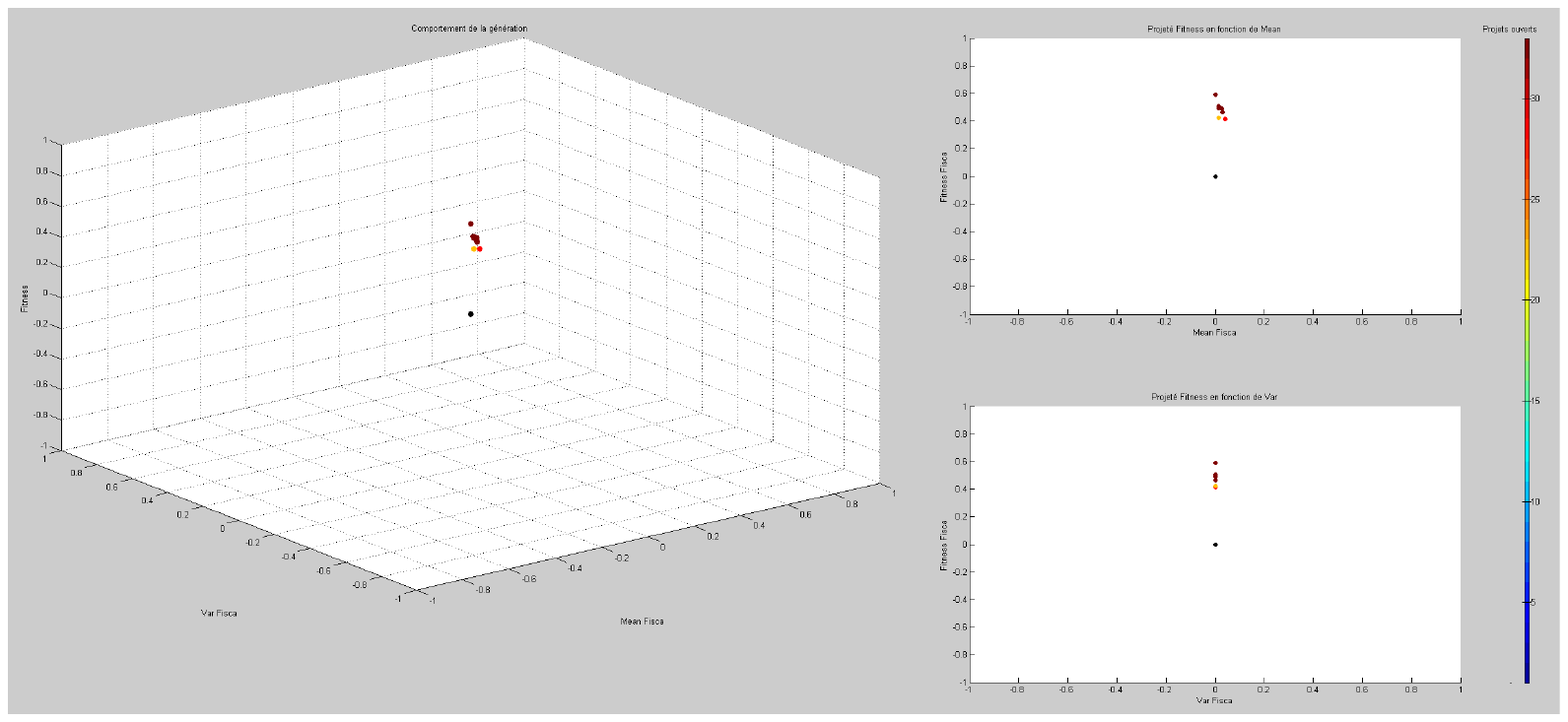}
\caption{The final population.}
\label{LabelFig10D5} 
\end{figure}

By constraining the solutions into a hypercube using the method described in this paper, we reach an optimum which has been validated by a finance professional as a reasonable solution for the community budget. The corresponding coding is:
\begin{gather}
\begin{aligned}
& \text {\tt [1,03949069282959, 0,19207769961155, 0,51186133809657, }
\\
& \text {\tt 0,205107769055541, 0,785367264162938, -0,254824609597842,}
\\
& \text {\tt 0,378497610225784, 1,04175330250962, 0,590217152232071,  }
\\
& \text {\tt -11,383992284572]}.
\end{aligned}
\end{gather}
These values correspond to:
\begin{gather}
\begin{aligned}
& \text{Project 1: OFF, Project 2: OFF, Project 3: ON, Project 4: OFF, Project 5: ON, }
\\
& \text{Tax evolution : 3.48\%, 2.85\%, 6.18\%, 3.02\%, 3.39\%.}
\end{aligned}
\end{gather}
The prudential ratios are respected, as the solution in the graphic simulator of Figure \ref{LabelFig10D4} shows (Capacity to Be Free of Debt, abbreviated as CDD in French, must remain under 15 years).
The fitness obtained is 59\%, and the 500 generations took 17 minutes and 32 seconds to be simulated on the reference machine. Interestingly, an independent computation with only 100 generations, which took 3 minutes and 32 seconds, showed a final fitness of 58\%, so the convergence is quite quick for this particular example. The corresponding chromosome was:
\begin{gather}
\begin{aligned}
& \text {\tt [1,11266759532518, 0,0838990704498006, 0,565754259647956, }
\\
& \text {\tt 0,440107396614116, 0,813652694225311, 0,642521321773529, }
\\
& \text {\tt 0,575349082741285, 0,447334636593206, 0,488786454202292, }
\\
& \text {\tt 0,990373758336907]}
\end{aligned}
\end{gather}
In terms of budget, this means:
\begin{gather}
\begin{aligned}
& \text{Project 1: OFF, Project 2: OFF, Project 3: ON, Project 4: OFF, Project 5: ON,}
\\
& \text{Tax evolution : 1.25\%, 1.53\%, 2.73\%, 6.88\%, 1.04\%}.
\end{aligned}
\end{gather}
One will notice that the activation of the different projects is the same as the other solution, whereas the choice of Tax Evolution pattern is quite different. 
A quick conclusion would be that the fitness is more dependent on the projects activation than on the tax evolution, but this would need a robustness analysis, 
and this was not the subject of the present study.
The most interesting part of the result is that, along the generations, the solutions found are quite concentrated, as shown in Figure \ref{LabelFig10D5}.
This should be compared to the initial Genetic Algorithm optimization without Gram Schmidt projection.
\begin{figure}[htbp]
~\hspace{-1.3cm}
\includegraphics[width=17.5cm, bb = 2.5cm 19.8cm 18.5cm 27.5cm, clip=true]{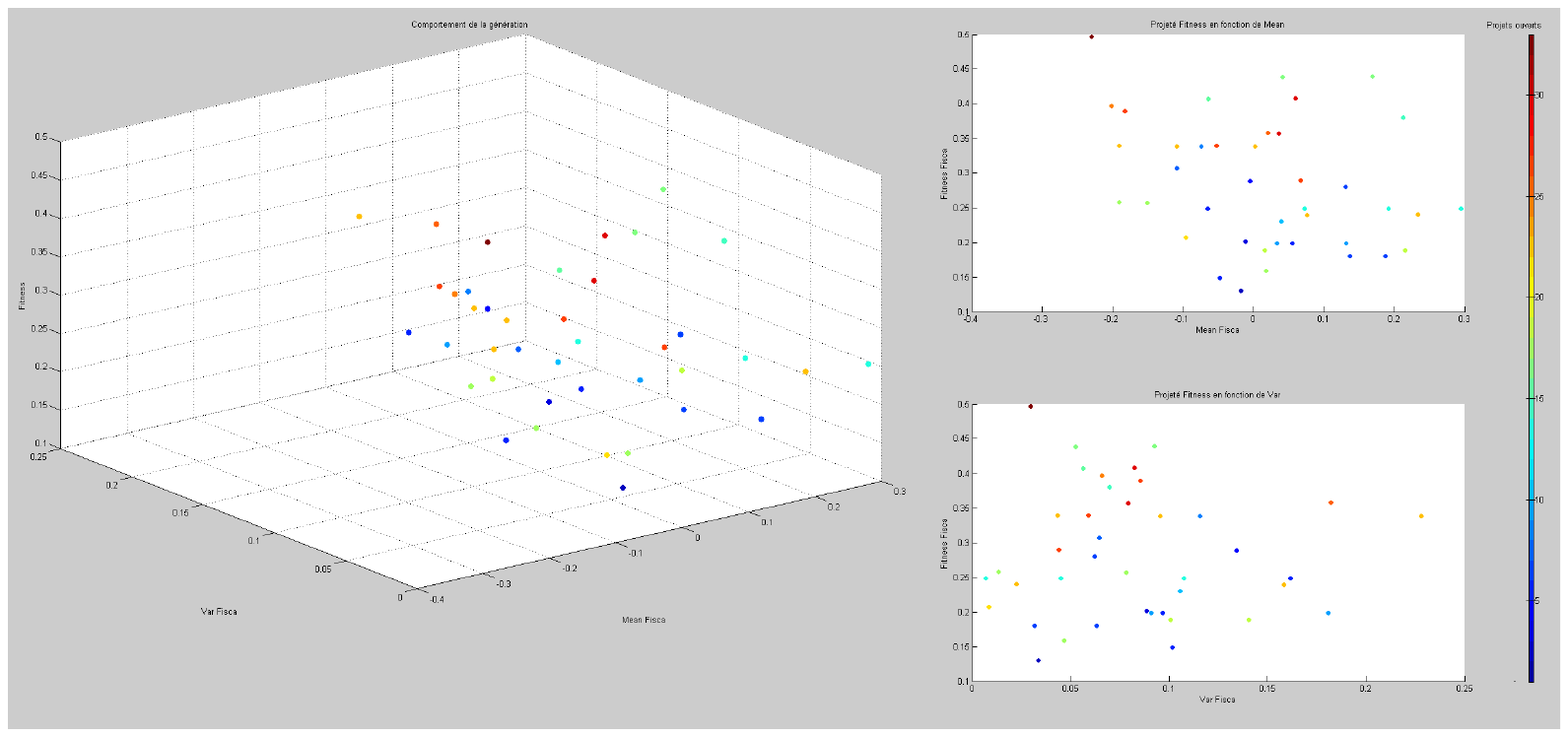}
\caption{The initial population.}
\label{LabelFig10D6} 
\end{figure}

The effect of using this particular technique is that the solutions are found into a constrained set of solution, without the user needing to explain how it is constrained, but by letting one propose two solutions surrounding the searched one.
The effect on the rapidity on the convergence was expected as an additional result of the study, but we could not demonstrate any noticeable or provable effect on this factor. Further studies need to be done, with a high volume of test, in order to determine whether the two points approach helps the convergence of the Genetic Algorithms or not, and depending on which conditions on the fitness of the coding of the chromosomes.
 \subsection{Parameter choices for the test}
 \label{SubsecPCCT} 
 Of course, the justification of the parameters chosen for the tuning of the Genetic Algorithms engine would make for an entire article on their own, 
 so we will just provide a justification for running tests like the one described in subsection \ref{SubsecTOP}.
More precisely, in this subsection, we report on a detailed study that allowed us to tune the parameters in this context.
 
The first part of the validation of values used for auto-shuffling, crossing, mutation and random selection rates are simply that they are the default values provided by the Open Source component AForge.Genetics. One can safely assume that these values have been chosen to be a correct fit to general situations.
 \\
 
The second part of the validation completes the first one, as the previous hypothesis has indeed been verified by the complete study, the outcome of which is shortly described below.
 
\begin{figure}[htbp]
\begin{center}
\includegraphics[width=9cm, bb = 0cm 0cm 14.8cm 5.9cm, clip=true]{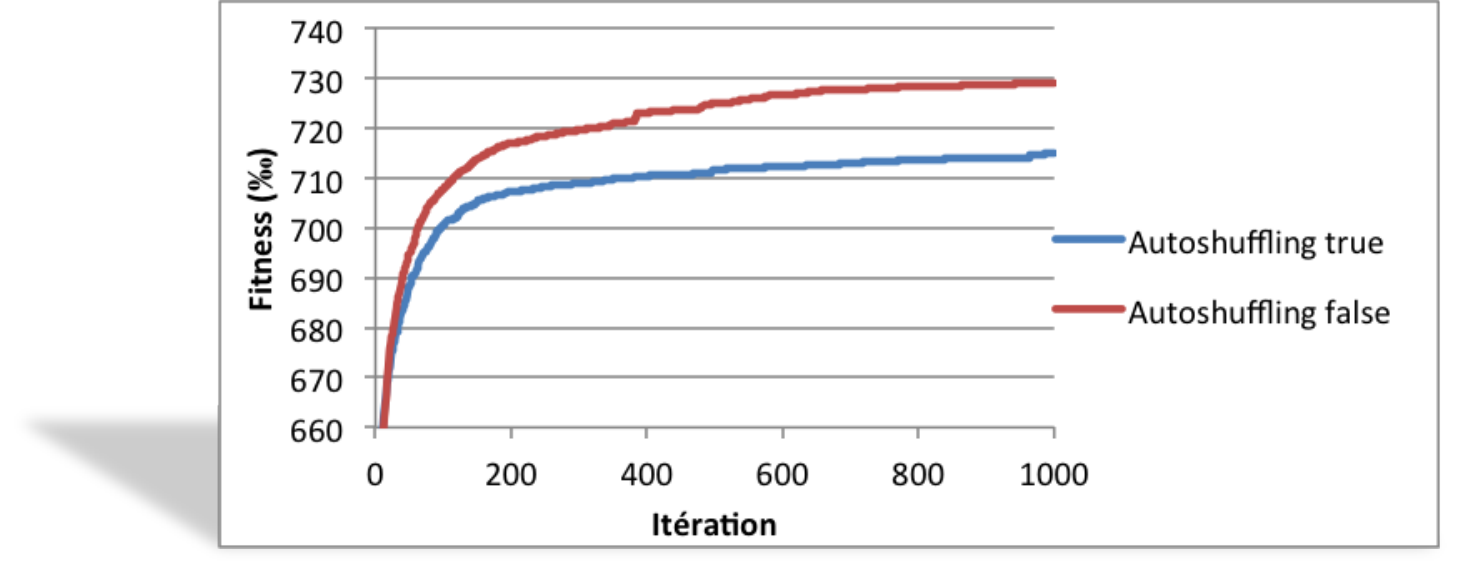}
\caption{Convergence with and without auto-shuffling.} \label{A-AutoShuffling} 
\end{center}
\end{figure}

\begin{figure}[htbp]
\begin{center}
\includegraphics[width=12cm, bb = 0cm 0cm 14.8cm 6.8cm, clip=true]{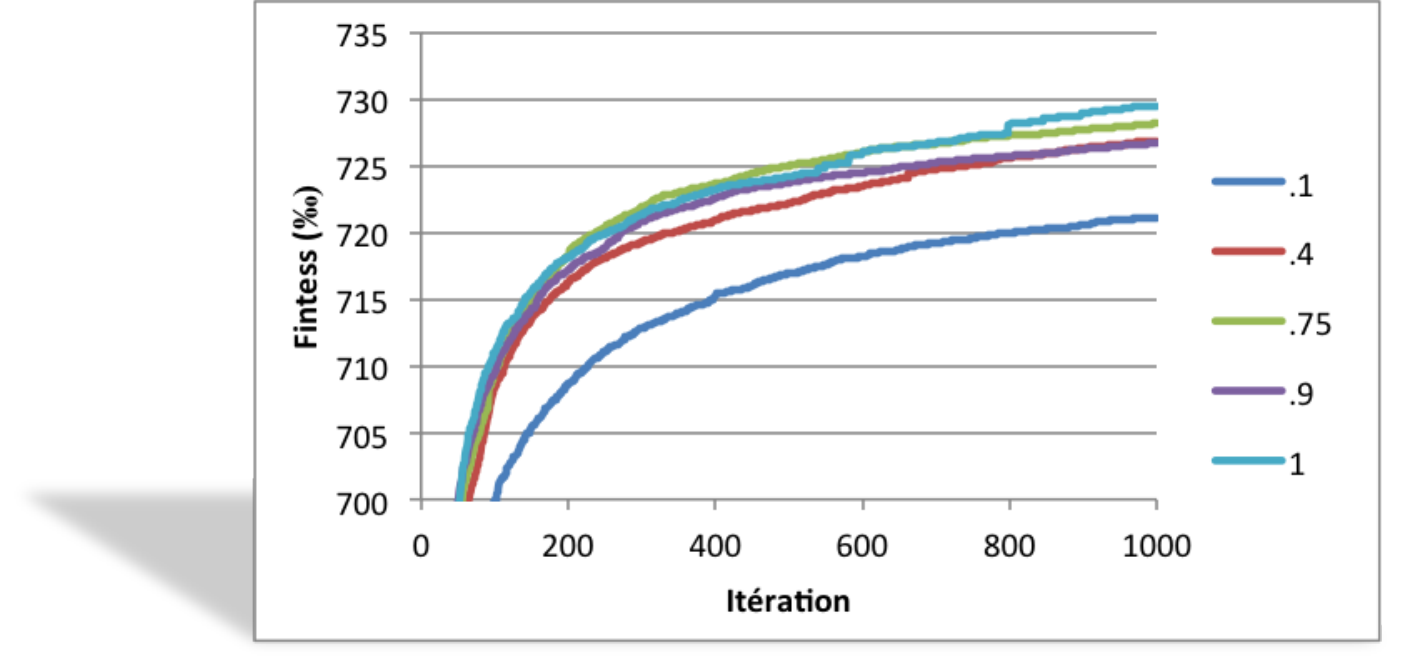}
\caption{Influence of crossing rate.} \label{B-CrossingRate} 
\end{center}
\end{figure}

\begin{figure}[htbp]
\begin{center}
\includegraphics[width=12cm, bb = 0cm 0cm 14.8cm 6.8cm, clip=true]{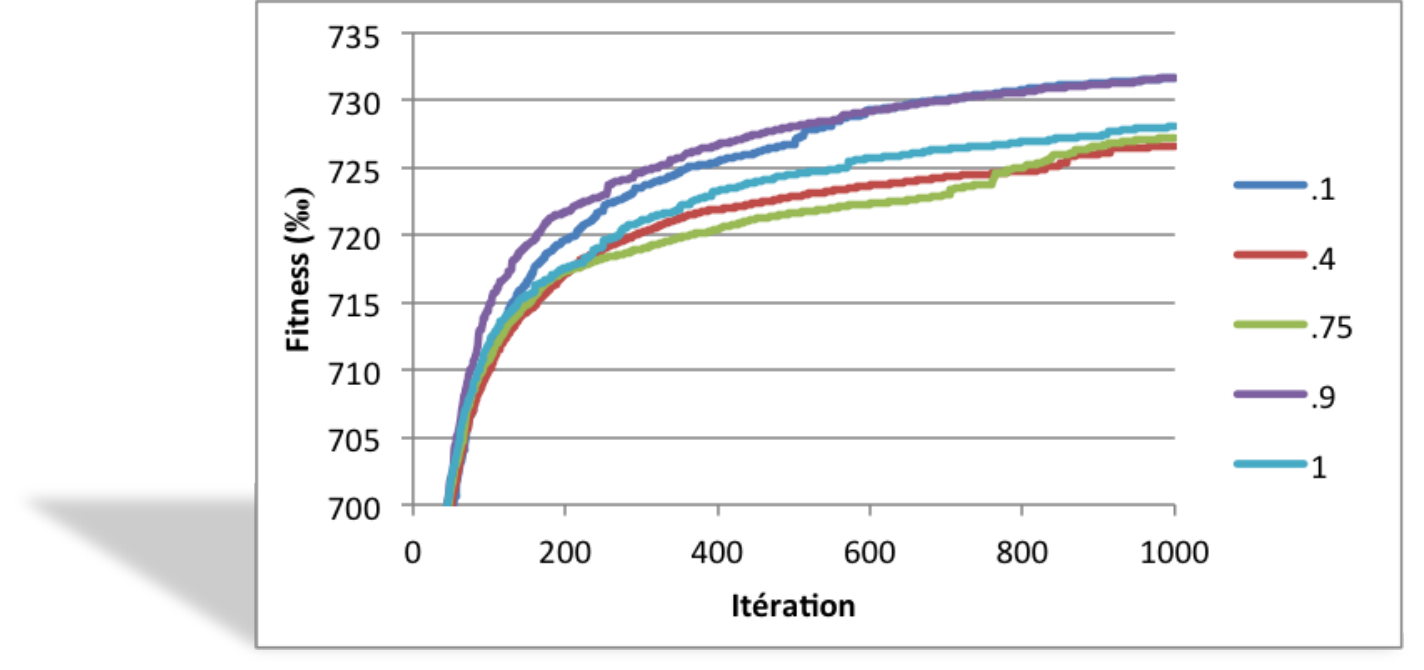}
\caption{Influence of mutation rate.} \label{C-MutationRate} 
\end{center}
\end{figure}

The auto-shuffling parameter allows for a dispersion of the chromosomes after selection. This is useful when the selection method forces the list 
 of upcoming chromosomes to be sorted, which can result in lower-quality crossing of the chromosomes thereafter. The "false" parameter 
 in our case is indeed the best in terms of convergence speed (external parameters like the size of the population have no impact on the 
 result) as shown in Figure \ref{A-AutoShuffling}.
The impact of the crossing rate has also been studied in the same conditions (accounting for variations of other parameters), 
and brought to the same conclusion that the default value of 0.75, provided by AForge, is quite optimal (see Figure  \ref{B-CrossingRate}).
Next analyzed parameter is the mutation rate, and again, the default value of 0.1 is close from ideal in our case of study 
(see Figure  \ref{C-MutationRate}).

\begin{figure}[htbp]
\begin{center}
\includegraphics[width=12cm, bb = 0cm 0cm 16.3cm 6.8cm, clip=true]{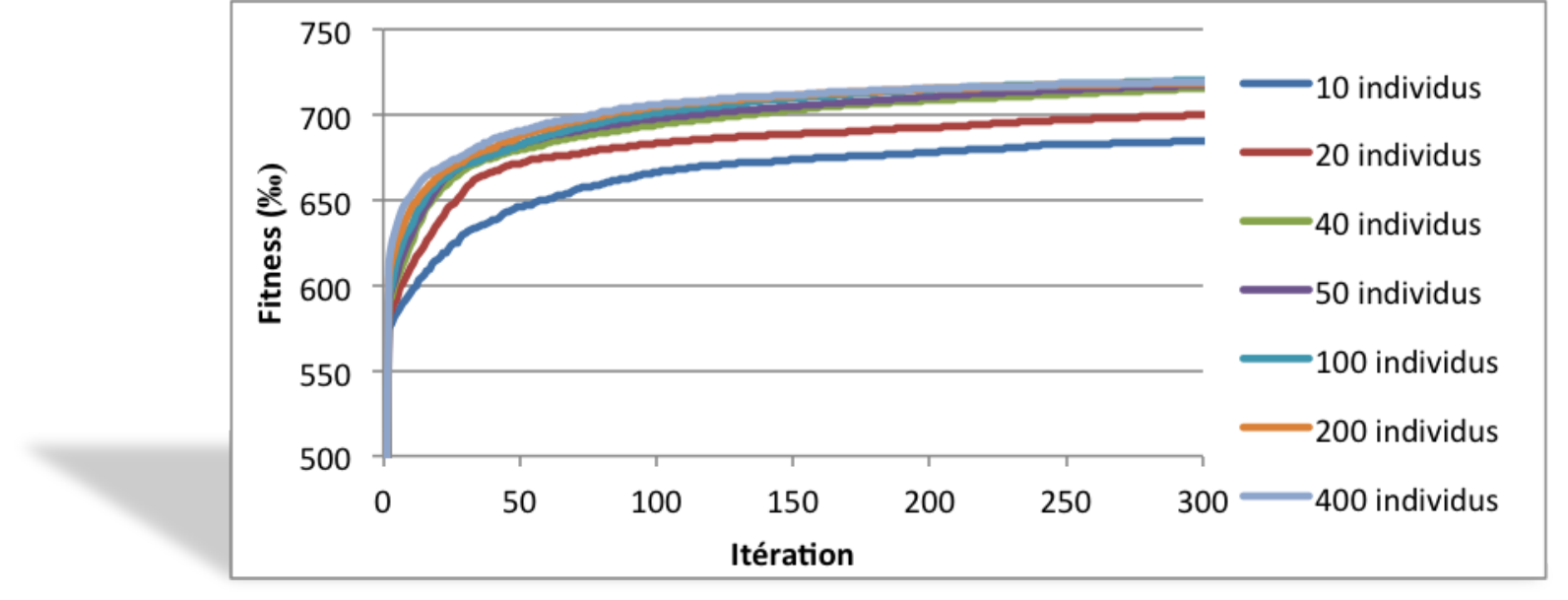}
\caption{Influence of population size.} \label{D-PopulationSize} 
\end{center}
\end{figure}

\noindent
The random selection rate is not used by the tailored selection algorithm created for the budget optimization engine. 
Finally, the impact of the size of the population on the convergence speed has also been analyzed (see Figure \ref{D-PopulationSize}), 
and brought us to use a value of 50, which was considered an optimal ratio between speed and memory use.
\\

As an aside note, the robustness tests around the four parameters needed to run several hundreds of hour-long tests, in order to simulate 
all the different possible combination. This has been achieved by creating a small software component that was dedicated in running 
jobs on colleagues' computers at night, collecting data and centralizing results back on the author's computer in the morning. 
This ad-hoc mechanism has also been used for the actual simulation, in order to test its robustness.
\\

Working with Genetic Algorithms makes for two competing risks that must be continuously balanced: under-fitting and over-fitting. In an under-fitting simulation, the generation algorithm introduces too much variety in the chromosomes for the fitness algorithm to restrain the population to a fitter one than the previous one. The over-fitting problem is the exact reverse, where the fitness algorithm completely stifles the expansion of the possible domain created by the generation algorithm, thus resulting in a lack of genetic variety, potentially leaving more ideal genes out of the simulation.
The under-fitting problem is head-on addressed by the very principle proposed in the current paper, which is to reduce the exploration domain to one restricted to the immediate neighborhood of two relatively satisfactory points. The restriction on the domain is achieved by choosing the chromosomes on constrained multi-dimensional boxes, the corresponding business values being then retrofitted in the standard domain of values by using the Gram-Schmidt routine.
The over-fitting problem is addressed by carefully choosing the selection method among commonly-accepted ones. On very continuous problems, the ''Elite'' selection mode achieves best results, by quickly removing the poor genes from the pool. On discontinuous problems like the ones linked to budget optimization, it is better to not be too harsh on the selection, and adopt a more exploratory selection mode, like "Roulette". In the "Roulette" method, the chromosomes are not simply eradicated if they do not correspond to a high ranking with respect to the fitness method, but have simply a lower chance of being selected for the next generation. This results in a more flexible approach, where the exploration of unknown domains is allowed, but more or less quickly forbidden if they do not bring an improvement on the fitness.
The question of the tuning between the "Elite" and the "Roulette" part of the selection algorithm could bear a complete study on its own. In the present study, this ratio has been chosen to a balanced default value of half / half, after that a considerable number of nightly robustness tests showed that increasing the exploratory part did not bring any better solution. After these tests, the ratio was kept for all subsequent simulations.
It would of course be possible to optimize the computation time by slowly decreasing this ratio to a more Elite-oriented algorithm, but the improvement on computational time (which cost was extremely low, all simulations being run on low-range PCs) would not make up for the risks on not detecting a better solution for the budget. This kind of adjustment is let for further study.

\section{Conclusion}
In this paper, a method based on Genetic Algorithm to build a collection of Financial Solutions from two acceptable ones is set out, and explored.\\
The way to tackle that the collection is sought in a neighborhood of the two acceptable solutions calls on a Gram-Schmidt routine to comfortably build a
box surrounding them. This routine brings also a way of coding the solution that can be used in the Genetic Like  Algorithm.\\
The method is then tested on simplified one-dimensional problems to exhibit that it has the capability to locate the argument of the maximum of a Fitness Function
and to generate a collection of solutions which is distributed over the set of all the arguments of the maximum when the Fitness Function has a plateau as a maximum.
This last capability is the important one in view of the targeted operational applications which concerns the financial strategy of local communities.\\
Finally, the method is tested on an example of the targeted operational applications and gives interesting results which is promising. It seems to be a potential alternative 
or a support to the heavy protocol (involving many meetings with experts and decision makers) to set out a suitable Financial Solution for  local communities.
\\

\noindent
{\bf Acknowledgements -} The authors thank the referee who pointed some lacks in the first version of the paper. 
He permitted a substantial improvement of the paper.

\bibliographystyle{plain}
\bibliography{biblio}

\end{document}